\documentclass[review]{elsarticle}
\usepackage{siunitx}
\usepackage{amsmath}
\usepackage{upgreek}
\usepackage{lineno,hyperref}
\usepackage{pdflscape}
\usepackage{afterpage}
\usepackage{longtable}
\usepackage{multicol}
\usepackage[width=\textwidth]{caption}
\modulolinenumbers[5]

\usepackage{setspace}
\usepackage{xcolor}
\newcommand{\red}[1]{\textcolor{black}{#1}}

\journal{Journal of \LaTeX\ Templates}

\begin{document}

\begin{frontmatter}

\title{Impacts on the Moon: analysis methods and size distribution of impactors.}

\author[oca,geo]{Chrysa Avdellidou\corref{mycorrespondingauthor}}
\cortext[mycorrespondingauthor]{Corresponding author}
\author[oca2,geo]{Edhah Munaibari}
\author[col]{Raven Larson}
\author[imcce]{Jeremie Vaubaillon}
\author[oca]{Marco Delbo}
\author[col]{Paul Hayne}
\author[oca]{Mark Wieczorek}
\author[abe]{Daniel Sheward}
\author[abe]{Antony Cook}

\address[oca]{Universit\'e C\^ote d'Azur, CNRS, Observatoire de la C\^ote d'Azur, Laboratoire Lagrange, Blvd de l'Observatoire, CS 34229, 06304 Nice cedex 4, France}
\address[geo]{Universit\'e C\^ote d'Azur, CNRS, Observatoire de la C\^ote d'Azur, G\'eoazur, 250 rue Albert Einstein, Sophia Antipolis 06560 Valbonne, France}
\address[oca2]{Master of Astrophysics - Universit\'e C\^ote d'Azur (MAUCA), D\'ept. de Physique (Campus Sciences, UCA) \& Observatoire de la C\^ote d'Azur, Parc Valrose, 06100 Nice, France}
\address[col]{Laboratory for Atmospheric and Space Physics, and Astrophysical \& Planetary Sciences Department, University of Colorado Boulder, Boulder, CO, USA}
\address[imcce]{Institut de M\'ecanique C\'eleste et de Calcul des Eph\'emerides, Observatoire de Paris/PSL, 77 Av Denfert Rochereau, F-75014 Paris, France}
\address[abe]{Department of Physics, Aberystwyth University, Ceredigion, United Kingdom, SY23 3BZ}

\begin{abstract}
We are preparing a telescope system to carry out a survey of detection and analysis of lunar impact flashes. 
In the framework of this project, here we present all necessary methods to automatically identify these luminous events, their lunar impact coordinates, the origin of the impacting meteoroids, as well as the estimation of their physical properties such as mass and size. We tested our methods against confirmed impact events and constructed the meteoroid size frequency distribution of impactors using literature data of the last 20 years. In addition, we present the first lunar impact event observed from the Observatoire de la C\^ote d'Azur that was detected during the testing phase of our project.
\end{abstract}

\begin{keyword}
 Moon, surface \sep Impact processes \sep Meteoroids
\MSC[2010] 00-01\sep  99-00
\end{keyword}

\end{frontmatter}


\section{Introduction}
\label{intro}

\red{Impact cratering is a fundamental process, altering planetary surfaces. Currently we have reached a level where we observe and study impact craters on a plethora of solar system bodies such as planets, moons, asteroids \citep{marchi2015, marchi2012, sugita2019, walsh2019}, Kuiper-belt objects \citep{singer2019} and \red{comet nuclei} \citep{vincent2015}. 
Cratering has been used, together with the knowledge of the size frequency distribution of potential impactors for a given collisional environment, in order to date planetary surfaces based on the simple idea that old surfaces have accumulated more impact craters than more recent ones. This started more than 50 years ago with the Moon \citep{baldwin1971} and was later used to study the chronology on other planetary surfaces such as Mercury \citep{neukum2001}, Venus \citep{korycansky2005}, and Mars \citep{ivanov2001,hartmann2001}. 
At the same time impact scaling laws have also been developed \citep[e.g.][]{housen1987, holsapple1994}.
Impact craters have been observed at different scales that span from kilometres, on planetary surfaces, down to microns, as for example on the returned lunar samples. NASA's sample return space mission OSIRIS-REx has recorded \emph{in situ} impact craters on the boulders of the target near-Earth asteroid (101955) Bennu, covering a range from 3~cm up to 5~m \citep{ballouz2020}.}

In parallel, laboratory experiments have advanced on the type of materials that are used to simulate planetary surfaces \citep{daly2015,daly2016, avdellidou2016,avdellidou2017,avdellidou2020, flynn2018,flynn2020}.
During laboratory experiments we are able to control and have accurate measurements of the mass, size and velocity of the impactor in order to study the crater formation. However, the size scales and velocity ranges (and thus the energy) are smaller than in planetary scales. On the other hand, at larger planetary scales cratering is mainly studied at a later time of the impact and there is no absolute link between the impacting body and the formed crater. \red{In this respect there have been executed several large-scale experiments of artificial impacts on planetary surfaces. Throughout the decades spacecraft on lunar orbits have eventually crashed on the Moon producing craters, several of which are identified and studied especially with the help of Lunar Reconnaissance Orbiter's (LRO) data \citep{plescia2016}. One of those anthropogenic impacts was ESA's SMART-1 \citep{foing2018} whose crater was detected years later in the LRO data \citep{stooke2019} and  NASA's LCROSS \citep{colaprete2010,schultz2010} whose impact was observed by the LRO/Diviner \citep{hayne2010} as well as from the Earth \citep{strycker2013}. On the nucleus of comet 9P/Tempel was performed the Deep Impact experiment by the Stardust-NExT mission that allowed the observation of the ejecta evolution at a real time \citep{schultz2007} but also later \citep{schultz2013}. In 2019 the Small Carry-on Impactor (SCI) of JAXA's Hayabusa2 sample return mission hit the surface of its target near-Earth asteroid (162173) Ryugu \citep{arakawa2020}, while a new large scale impact experiment is being prepared on Dimorphos, the satellite of near-Earth asteroid (65803) Didymos, by NASA's DART mission in 2022 \citep{cheng2016}. In several cases laboratory impact studies have been done in order to accompany the aforementioned artificial experiments \citep[see for example][]{schultz2005, burchell2015}. }

In this respect, since late 90s, the lunar surface has been monitored in order to detect transient light phenomena that occur during meteoroid hypervelocity impact events, and we refer to these events as lunar impact flashes (LIF). These observations have been performed by small and moderate telescopes from the ground with \red{the} aim to derive the impact flux of cm- to dm-sized meteoroids \citep{suggs2014,larbi2015,bonanos2018,avdellidou2019}. In parallel, several detection and analysis techniques have been developed depending on the specific observational setup (single or double camera systems) and the scientific goals of the teams \citep{bouley2012,madiedo2015,xilouris2018}. In a few cases NASA's Lunar Reconnaissance Orbiter (LRO) has detected these impact craters on the lunar surface \citep{suggs2014,robinson2015,ortiz2015}.

In this work we use current techniques of lunar observations and develop a methodology to observe and study live impacts of \red{cm to dm-size} impactors in space. \red{For our project the surface of the Moon provides an excellent opportunity because it is the closest planetary body to Earth; it suffers impacts from known populations of cm to dm-size objects (when they are linked to meteoroid streams, but less certain for the sporadic meteoroids); lunar orbiters provide data for crater detection; and low-cost instrumentation can be used worldwide for a continuous monitoring of the lunar surface. Additional advantage of observing impacts on an atmosphere-less body such as the Moon is that the impacting object is allowed to reach the lunar surface without breaking. In that way we could estimate its original size and establish a link between the impacting meteoroid and the produced crater. At the same time meteoroid impacts on planetary objects with atmosphere provide complementary advantages such as the estimation of their strength and density, the latter crucial parameter for size estimation techniques, as will also be presented later in this work. The lunar impactors we intend to study here have small sizes and velocities that, if originate from meteoroid streams, can reach up to 70~km~s$^{-1}$, while at their lower end, coming from asteroid orbits \cite{bottke2002}, their impact velocity distribution has a maximum at about 12~km~s$^{-1}$ allowing velocities down to $\sim$5~km~s$^{-1}$ \cite{lefeuvre2011}. In a broader picture we can consider those impacts are an extension in size and energy scales of the current laboratory impact experiments using light-gas guns, where masses are up to a few grams and velocities limited to $\sim$8~km~s$^{-1}$.}

We are building a new observational collaborative survey in order to observe the lunar surface starting from two sites, in France and the UK. Our scientific goal is to perform statistics of the meteoroids regarding their size frequency distribution and subsequently detect their craters, establishing a link between the size of the crater and the impactor. In addition, we want to study differences in physical properties between the different population of the impactors (different meteoroid streams which are known to have different composition and material density) as well as the cooling rate of the molten lunar regolith, hunting for differences between impacts on maria vs. highlands. In those two lunar soils the grain size of the regolith is the same, but the mineralogy differs from basalt to anorthosite respectively.

We collected LIFs from the literature, which are used for two purposes: (i) as input to build our LIF analysis methods and (ii) to construct the size frequency distribution of small meteoroids that hit the Moon. In addition, we have performed our own test observations and built our custom LIF detection algorithm. 
Specifically in $\S$~\ref{data} we present the datasets that we used for our analysis, in $\S$~\ref{detection} we describe our methods to identify potential LIFs in observational data. In $\S$~\ref{coords} we describe in detail how we derive the selenographic coordinates of the events and in $\S$~\ref{source} how we identify the origin of each meteoroid impactor. The estimation of the mass and size of the impactors is presented in $\S$~\ref{masses}. In $\S$~\ref{flash_oca} we report our first detection that was observed during the testing phase of the project and finally in $\S$~\ref{discussion} we discuss our results.

\section{Data}
\label{data}
\red{For the purposes of this work we use the publicly available data that were acquired in the last two decades and obtained from scientific publications or official websites. More specifically, we tested the performance of our algorithms on the NELIOTA dataset, while all the datasets presented bellow were used to derive the size frequency distribution of the impacting meteoroids.}

\subsection{NELIOTA}
The ESA-funded program NELIOTA run at the National Observatory of Athens with a goal to detect LIFs. As of December 1$^{st}$ 2020 there have been detected, by NELIOTA's double camera system, 112 confirmed LIFs in about 167 hours of observations since February 1$^{st}$ 2017, giving a rate of 1 flash every 1.5 hours of observations. Extensive details on the instrumentation, the observational method and calibration techniques are given by \cite{xilouris2018} and \cite{bonanos2018}. 
\red{The} NELIOTA website\footnote{neliota.astro.noa.gr} provides data for the date and time of each event (Tab.~\ref{table3}), selenographic coordinates (Tab.~\ref{app_table1}), duration and the calibrated peak magnitudes in $R$ and $I$ bands. Unfortunately, as it is derived by \cite{xilouris2018} and \cite{liakos2020}, the reported magnitudes and errors on the NELIOTA database are approximate, while some selenographic coordinates are possibly reported incorrectly (swapped longitude and latitude). In this work, for the flashes with numbers 1-80 we use the published magnitudes and selenographic coordinates \citep{xilouris2018, liakos2020}, while for the rest (81-112) we use the data as reported on the website. As we will see later, we use our method and re-compute the selenographic coordinates (Tab.~\ref{app_table1}), which we use for the rest of the analysis.
Additionally, we retrieved the datacubes that include the total frames in which a flash appears. These were used in order to estimate the $R$ and $I$ magnitudes of the subsequent frames in the multi-frame events \citep{avdellidou2019}.

\subsection{Marshall Space Flight Center}
Since 2006, NASA's Marshall Space Flight Center (MSFC), has started an observational program, which monitors the lunar surface, hunting for LIFs. Observations are conducted at NASA Marshall Space Flight Center in Huntsville, Alabama at the Automated Lunar and Meteor Observatory (ALaMO). Detailed description on their instrumentation and observational methods can be found in \cite{suggs2014}. 
Until 2018 435 LIF detections had been published online\footnote{https://www.nasa.gov/centers/marshall/news/lunar/lunar\_impacts.html}, however, in this work we will use only the published data (126 events) by \cite{suggs2014}. This dataset contains information about the date, time, solar longitude, $R$ magnitude, selenographic coordinates (Tab.~\ref{app_table1}) and luminous energy (Tab.~\ref{table3}). The reason for which we did not use the full online dataset is that no magnitude information is provided, and thus no further analysis could be done for our purposes.

\subsection{Other observations}
Apart from the two aforementioned surveys there are numerous LIF observations that were done by different teams in Spain \citep{ortiz2002, ortiz2006,ortiz2015,madiedo2015b, madiedo2015, madiedo2016, madiedo2018}, Japan \citep{yanagisawa2002,yanagisawa2006}, USA \citep{dunham2000, cudnik2003}, Morocco \citep{larbi2015} and Mexico \citep{ortiz2000} using small telescopes with diameters between 20 and 40~cm. From all these studies we have collected the date and time (Tab.~\ref{table3}), impact duration, magnitude and selenographic coordinates for 69 events in total (Tab.~\ref{table3}). The luminous energy was given only for five events \citep{larbi2015}.

\section{Detection of events}
\label{detection}
The first step is to identify all the potential LIF events in the datasets of the observations.
We developed two approaches that can be used to carry out real time detection of LIFs. Both approaches are based on the creation of a master reference frame, which we call the Lunar Background (LB), that is subtracted from all frames of the night side of the Moon. This is necessary in order to remove the inhomogeneous surface of the Moon and the earthshine. From each frame we subtract a LB that is created by the median combination of the 10 previous frames and thus any potential luminous transient phenomenon should remain in the images. \red{This analysis step will be carried out synchronously with the observations and the data acquisition of our developing observing system.}

In the first detecting method we apply a threshold on the pixel values of the background-subtracted frames. Then, to ensure that all the pixels detected to be above the threshold are not due to any artefacts, we developed and applied multiple filters. This is to confirm that the pixels of a LIF form a Gaussian shape.  
 As this method is based on the application of a threshold, a careful selection of the values of those criteria is required, as larger values can result on missing fainter impacts and lower values may increase the number of detected artefacts (false detection).

The second method, is to apply a Gaussian filter to the background-subtracted images. This will smooth most of the artefacts leading to their elimination. In this case we do not need to apply cleaning filters (as in the first approach) saving computational time. This process will enhance the pixels of a potential event which leads to an easy detection even for faint LIFs. In that way we overcome the issue of incorrect selection of any threshold, as described in the first method.

The most significant consideration when developing these real-time detection methods was the fast execution and the short computational time in order to keep up with the running observations. Both methods were tested against data containing synthetic LIFs that we generated, \red{as well as against the archived data from NELIOTA}. The results of our examination show that both approaches can keep up with real time observations, detect all events (real or simulated) with the first approach being faster but less sensitive to fainter LIFs.

\section{Selenographic Coordinates}
\label{coords}
The importance of accurately locating the selenographic latitude and longitude ($\beta,\,\lambda$) of a LIF comes from the fact that it is needed to perform the process of linking the impactor to its source. Moreover, it allows us to investigate the aftermath of an impact, by searching for the produced crater or the fresh ejecta deposits using spacecraft data from the LRO \citep{sheward2019}. 

As mentioned before, the selenographic coordinates of the confirmed LIFs are also reported in the literature (Tab.~\ref{app_table1}). For the sake of our preliminary survey, we have developed the Automated Geolocation of Lunar Surface Impacts (AUGUR) algorithm to identify the selenographic coordinates from individual observations \citep{larson2019}. In order to test AUGUR's performance we applied it to the NELIOTA events, as it is the only survey that provides \red{access to the complete datacube of each detection, in the form of FITS files.}

\begin{figure}
	\centering
	\includegraphics[height=15cm]{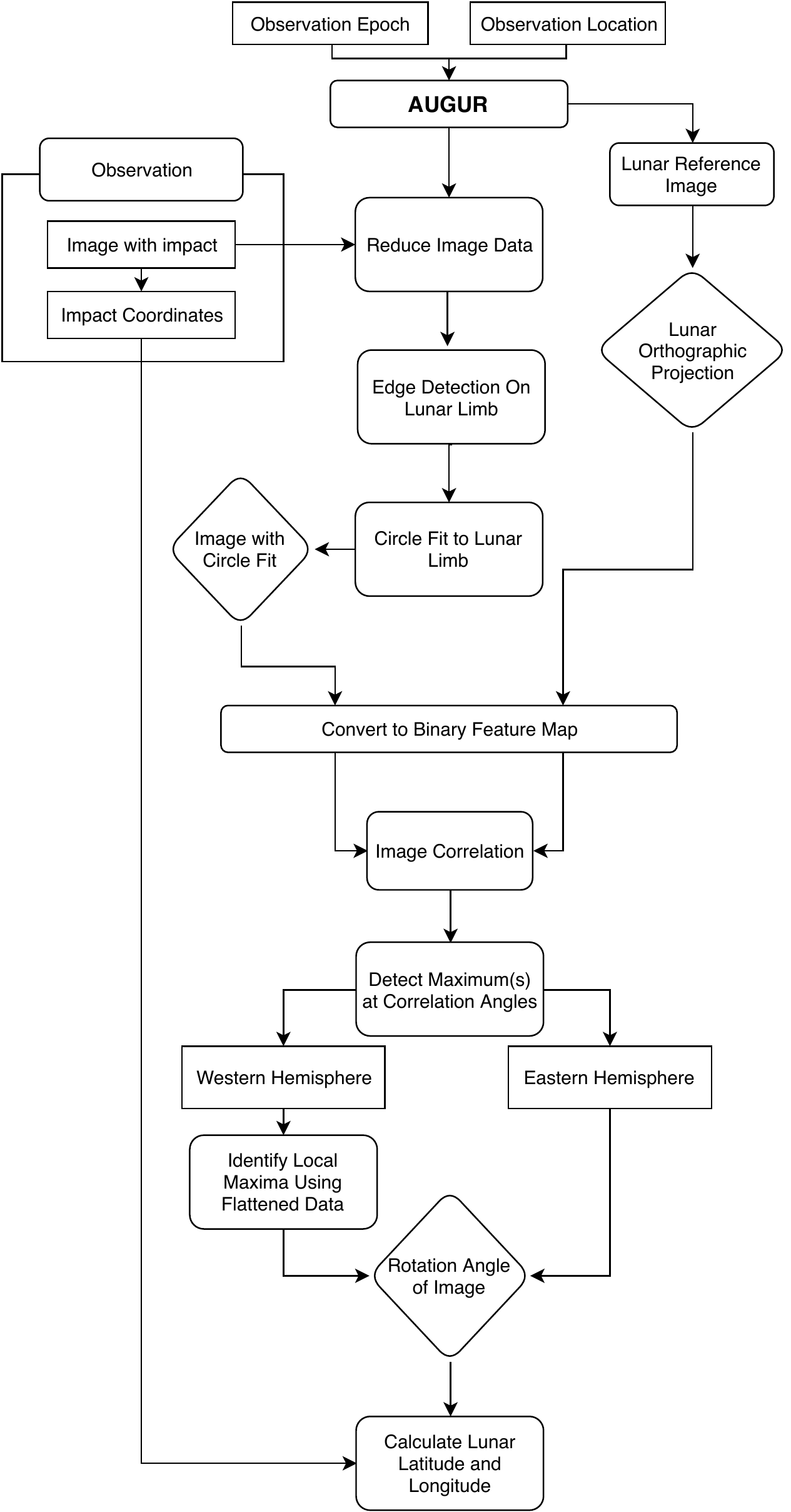}
	\caption{The full process of the AUGUR algorithm. Rectangular boxes are inputs to the program and diamond boxes are outputs.}
	\label{fig:flowchart}
\end{figure}

The AUGUR algorithm takes in images, containing a LIF, and returns the selenographic coordinates of that impact and its respective error. Once an impact flash is detected in an image, AUGUR can be automatically triggered to geolocate the selenographic coordinates of that flash. Fig.~\ref{fig:flowchart} shows the full mechanics of the AUGUR algorithm, however, this will be broken down into the three main steps of image reduction, image correlation, and coordinate transformation. 

Before AUGUR begins the image reduction and correlation, it first uses the \textit{JPL Horizons} database to pull information to generate a projection of the Moon for each observation. This projection uses an orthographic projection matrix and takes into account lunar librations by setting the sub-observer latitude and longitude as the (0,0) coordinate for the projection. The projection is wrapped with the \textit{Moon LRO LROC WAC Global Morphology Mosaic} to build a projected Moon that can later be used by AUGUR during image correlation. AUGUR performs best when provided a datacube of a minimum 10 lunar images, which should contain the lunar limb. Terminator could be contained as well. 

\begin{figure}
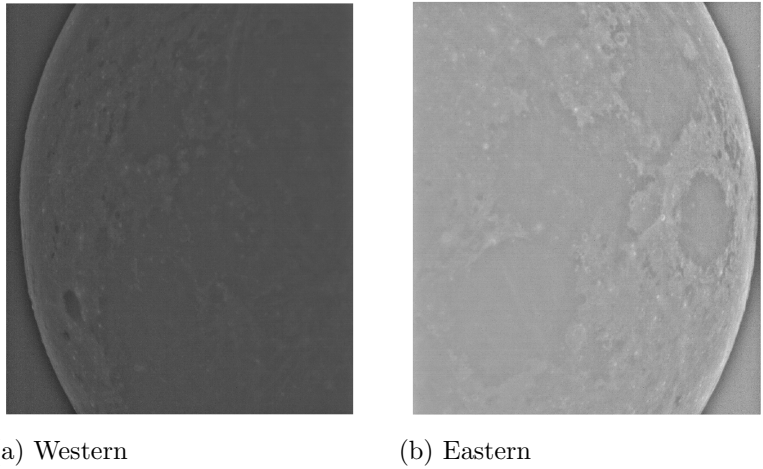

	\centering
	\begin{tabular}{ll}
		\includegraphics[width=5cm]{fig2a.pdf} &   \includegraphics[width=5cm]{fig2b.pdf} \\
		(a) Western & (b) Eastern \\
	\end{tabular}
	\caption{Results of the stacked image being sent through a low-pass filter to remove the light gradient but preserve the lunar features. Both of these images were taken at an illumination of $21\% - 22\%$ resulting \red{in} good feature preservation after the light gradient was removed. Images taken from the NELIOTA website.}
	\label{fig:gradrmv}
\end{figure}

The first step of AUGUR is to reduce the image data. Once the datacube is loaded into AUGUR, it is essential to stack the images to bring out as many lunar features as possible. Since the images are taken using a high speed imaging system, distortion in between frames is negligible and they can be stacked without temporal alignment. After the images are stacked, it is then necessary to remove the light gradient coming from the terminator of the Moon or from any Earthshine presented in the image. This can be done by putting the stacked image through a low-pass filter, thus preserving the lunar features but removing the saturation of excess light in the images. For a low-pass filter AUGUR applies a large Gaussian blur ($\sigma = 25$ pxl) to the stacked image by calculating a Gaussian filter kernel using 

\begin{equation}
G(x,y)=\frac{1}{2\pi\sigma^2}e^{-\frac{x^2+y^2}{2 \sigma^2}}
\label{eq:gaussianfilter}
\end{equation}
where $\sigma$ is the standard deviation of the Gaussian, and $(x,y)$ are the pixel coordinates of the image. We can then convolve that kernel with the original stacked image using

\begin{equation}
V =  \left | \frac{\sum_{x=0}^{\sigma} (\sum_{y=0}^{\sigma} G_{xy}d_{xy})} {F}  \right |
\label{eq:convolution}
\end{equation}
where  $d_{xy}$ is the data value of the pixel, and $F$ is the sum of the coefficients within the kernel (this is set to 1 if the sum of the coefficients is 0). Once the Gaussian blur is applied to the stacked image, the light gradient is all that remains. To remove the light gradient from the stacked image, we divide the stacked image by the blurred image to leave only the lunar features. The result of this low-pass filter can be seen in Fig.~\ref{fig:gradrmv}.

The next step is to automate the detection of the lunar limb in the reduced images. This is later needed to align the image with the projected Moon and correlate where the image is located within the lunar hemisphere. To detect the lunar limb, the image is run through a Sobel filter to enhance the lunar limb from the background. The Sobel filter method is similar to the low-pass filter method laid out above in that it passes the image through a Gaussian filter and then convolves that kernel with the original image. However, the Sobel filter requires a smaller Gaussian blur effect and uses a $3 \times 3$ ($\sigma$ = 3~pixels) kernel. 

The Sobel edge detection uses two convolution kernels ($G_x, G_y$) with one detecting change in the x-direction of the image and the other detecting change in the y-direction of the image. The $G_x$ kernel is simply the $G_y$ kernel rotated by 90 degrees. To detect the lunar limb, the magnitudes of the kernels are calculated using 

\begin{equation}
|G_{mag}| = \sqrt{G_x^2+G_y^2}.
\label{eq:sobel}
\end{equation}

\begin{figure}
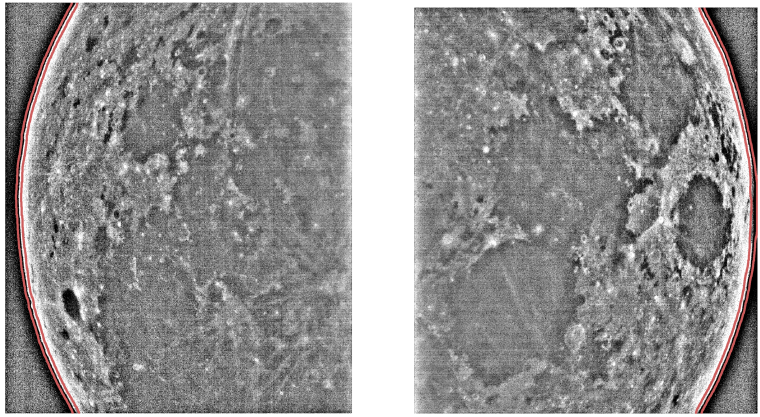

	\centering
	\begin{tabular}{ll}
		\includegraphics[width=5cm]{fig3a.pdf} &   \includegraphics[width=5cm]{fig3b.pdf} \\
		(a) Western & (b) Eastern \\
	\end{tabular}
	\caption{\red{Light gradient removed images with the inner and out edge of the lunar limb detected (red). The average of these two detections is used for the final location of the lunar limb, resulting in $\sim$1 pixel accuracy.} These images do not represent the magnitudes of the Sobel filter but are instead the light gradient removed images with a pixel threshold applied to them for contrast. Images taken from the NELIOTA website.}
	\label{fig:gradrmvedg}
\end{figure}

The higher resulting magnitudes correspond with a larger gradient change in the pixel grid. AUGUR then locates pixels that correspond to higher magnitudes and saves them as the potential $x$ and $y$ coordinate pairs of the lunar limb, as seen in Fig.~\ref{fig:gradrmvedg}.

After the detection of the lunar limb is complete \red{(with $\sim$1~pxl accuracy)}, a circle can be fit to the limb with a least squares circle fit method using 

\begin{subequations}
	\begin{equation}
	\label{eq-a2}
	g(u,v) = (u-u_c)^2 + (v-v_c)^2 - \alpha
	\end{equation}
and
	
	\begin{equation}
	\label{eq-b2}
	S = \sum_i(g(u_i,v_i))^2
	\end{equation}
\end{subequations}
where $\alpha = R^2$ and $u_c, v_c = (x_c,y_c) - (\bar{x}, \bar{y})$ provides the central coordinates of the best fit circle in the original ($x,y$) coordinate system of the image, assuming $u_i = x_i - \bar{x}$, and $v_i = y_i - \bar{y}$.

\begin{figure}
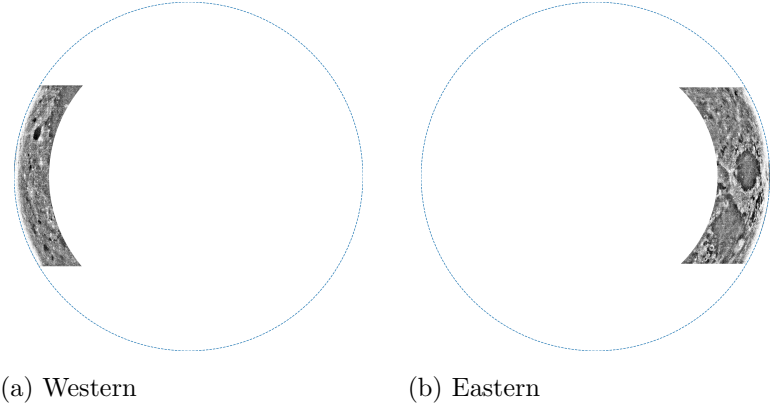

	\centering
	\begin{tabular}{ll}
		\includegraphics[width=5cm]{fig4a.pdf} &   \includegraphics[width=5cm]{fig4b.pdf} \\
		(a) Western & (b) Eastern \\
	\end{tabular}
	\caption{The masked images with their corresponding least-squares circle fit. Images in the Western of the Moon are masked to show $20\%$ of the image, and images on the Eastern hemisphere are masked to show $30\%$. \red{This is to deter AUGUR from a false rotation angle detection by focusing near the lunar limb where the contrast for the lunar features is more prominent.} AUGUR best identifies Grimaldi crater for the Western hemisphere and Mare Crisium for the Eastern. Images taken from the NELIOTA website.}
	
		\label{fig:circmnt}
\end{figure}

Once the central coordinates and radius of the best fit circle are found, AUGUR masks the image so that only the portion of the image close to the limb is visible in the circle fit (Fig.~\ref{fig:circmnt}), this allows for the image correlation to focus on certain lunar features in the image. We use a mask on the inner portion of the data to cancel out any \red{effect} the light gradient had on the lunar features, which would lead to a false correlation of the image later. 

Computing a least-squares circle fit for the Sobel filter data allows for us to align the lunar limb in the image exactly with the projection of the Moon generated for each observation. Locating the lunar limb constrains the image data to a certain part of the lunar surface, however the rotation of that image compared to the sub-observer latitude and longitude is unknown for each data set. Using the circle fit allows for us to correlate the image among different rotation angles by rotating it around that common central axis with the lunar projection.

Before comparing the image to the lunar projection, we first convert both the image and the projection into a binary feature map by using image thresholding. Using a pixel threshold brings out notable lunar features such as Grimaldi crater and Mare Crisium for AUGUR to focus on during the correlation. The thresholding limit assumes a darker pixel value corresponds to a more notable lunar features and patterns (such as the lunar Mare) and assigns those with a value of 1 in the binary map. 

\begin{figure*}
	\centering
	\begin{tabular}{cc}
		\includegraphics[width=50mm]{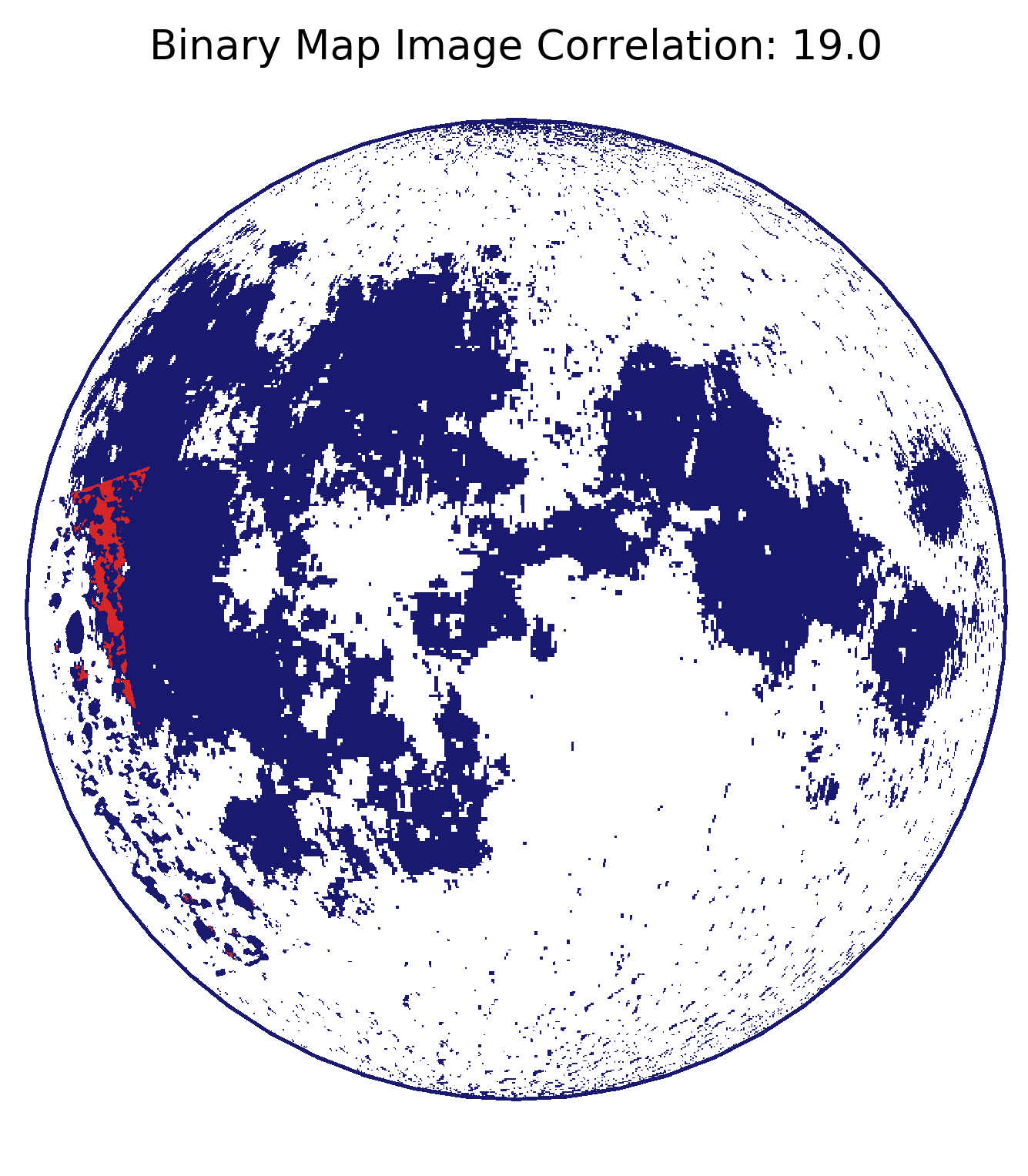} &
		\includegraphics[width=50mm]{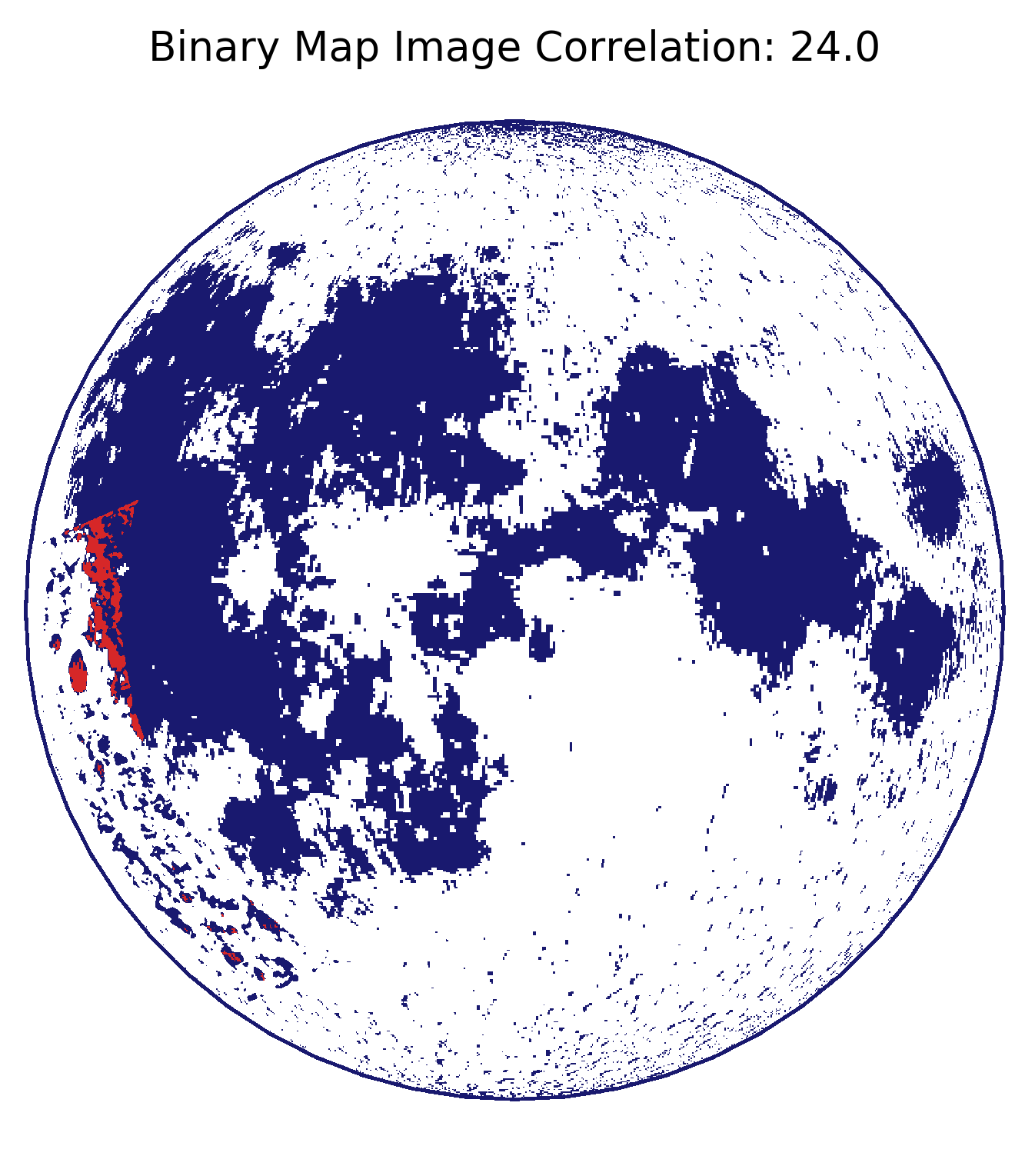} \\
		\includegraphics[width=50mm]{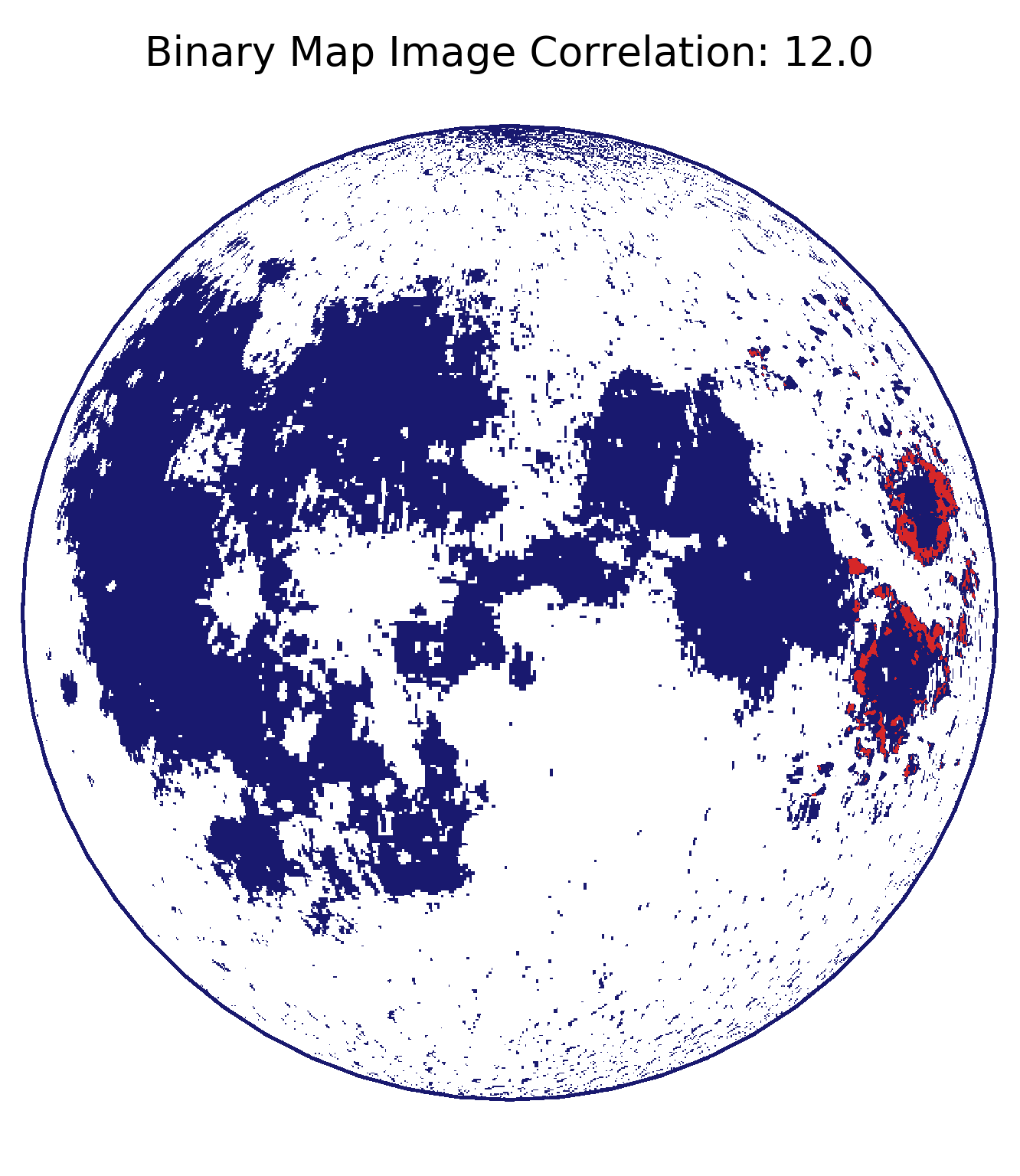} &
		\includegraphics[width=50mm]{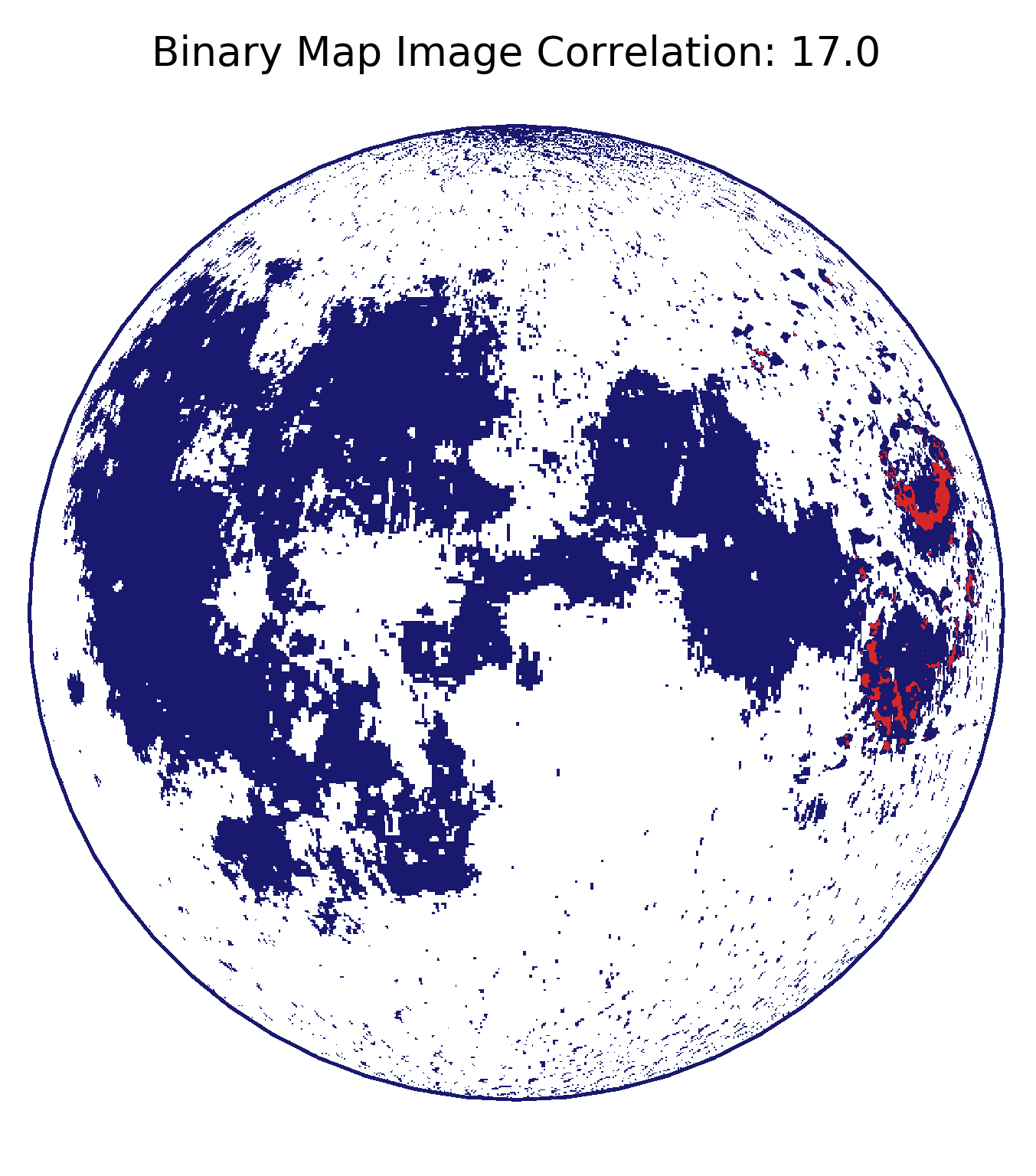}  \\
	\end{tabular}
	\caption{\red{A binary version of the masked images in Fig.~\ref{fig:circmnt} is  compared to the lunar projection threshold image.} The white background corresponds to no features, the dark blue relates to a feature in one image but not in the other, and the red pixels represent a positive match between images. The top progression shows a typical correlation match for the Western hemisphere and the bottom shows one for the Eastern hemisphere. \red{The number in each panel corresponds to the rotation degree of the image that is being tested.} The central image shows the correlation at the correct degree (this was found to be 12 degrees for the Eastern image and 24 degrees for the Western). The surrounding images show examples of correlation matches for incorrect rotation degrees.}
	\label{fig:binary}
\end{figure*}

\begin{figure*}
	\centering
	\begin{tabular}{cc}
		\includegraphics[width=6cm]{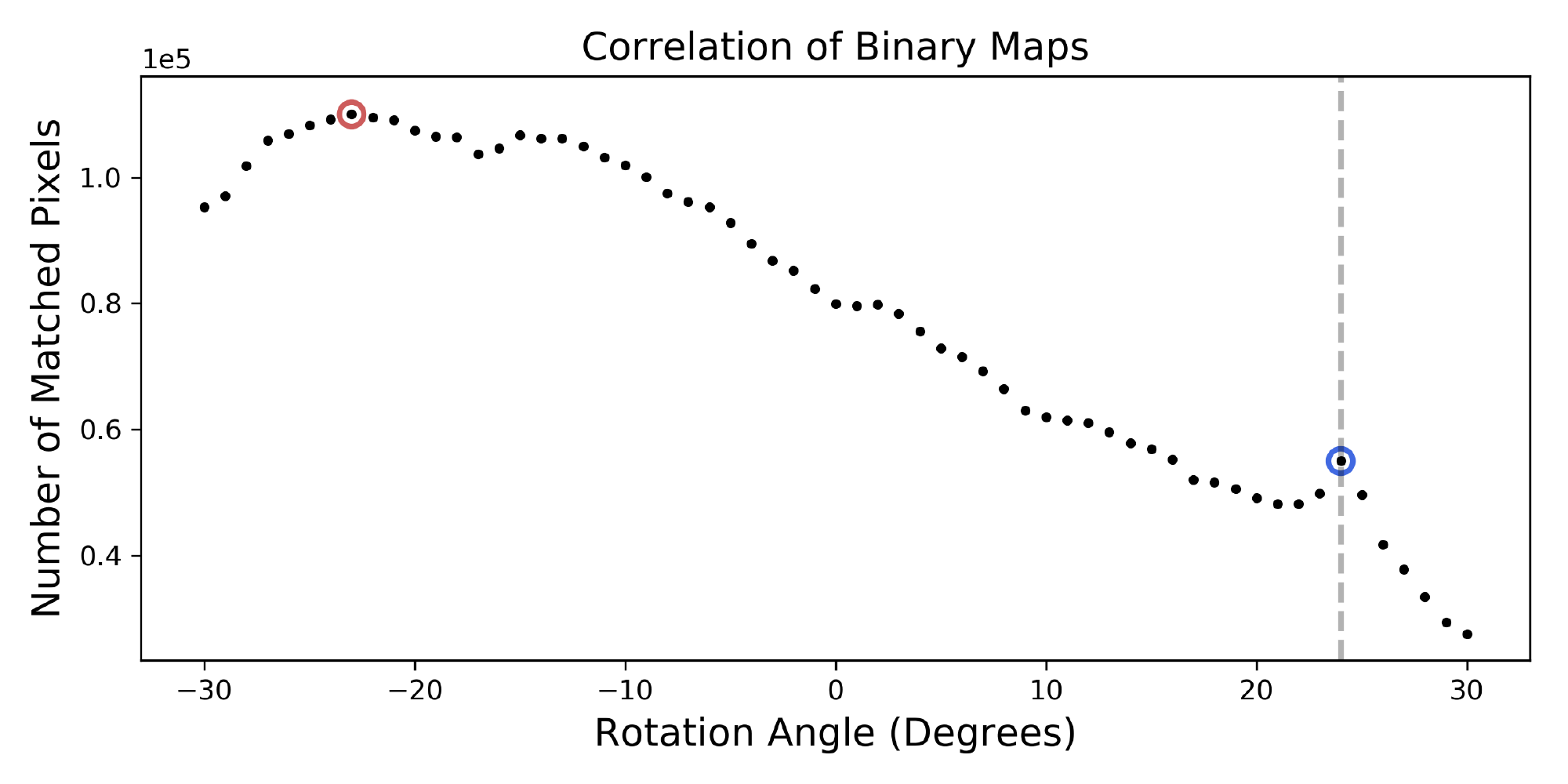} & \includegraphics[width=6cm]{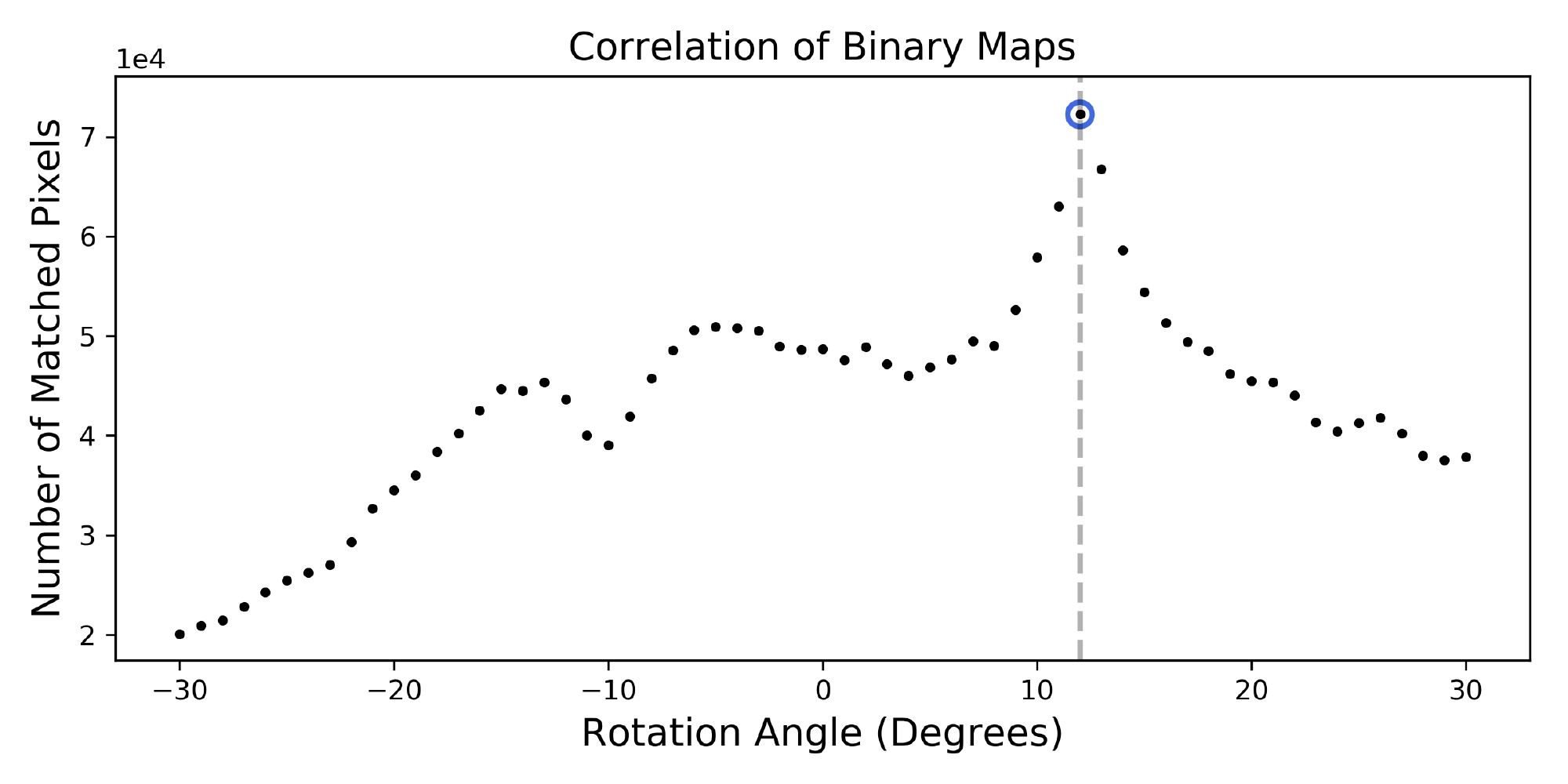} \\
		(a) Western & (b) Eastern \\[6pt]
	\end{tabular}
	\caption{Results of the correlation between the two binary images for each hemisphere. The x-axis represents the rotation degree of the image and the y-axis corresponds to the number of pixels matched between the images at that degree. \red{The blue circle and vertical dashed line indicates the correct rotation degree, where the red circle indicates a possible false rotation degree match. (a) The global maximum of the data does not correspond to the correct rotation angle and produces a false positive match, thus needs further analysis and (b) shows the same process for the Eastern hemisphere, however, the maximum matches with the correct rotation angle and does not need further analysis.}}
	\label{fig:peakdtc}
\end{figure*}

Once AUGUR converts both the impact image and the lunar projection into binary feature maps, it is then possible to measure the correlation at different rotation angles by overlaying the two binary images and counting the number of pixels that match in each. AUGUR counts the number of matched pixels for rotation angles between $\pm30$ degrees, with a 0 degree rotation angle starting at due East (0 degrees) for the Eastern hemisphere or due West (180 degrees) for the Western hemisphere. The positive direction for the rotation degree angles follows a counter-clockwise direction. Binary map comparisons for both hemispheres can be found in Fig.~\ref{fig:binary} with the correlation results in Fig.~\ref{fig:peakdtc}.

\begin{figure}
	\centering
	\begin{tabular}{cc}
	\includegraphics[width=6cm]{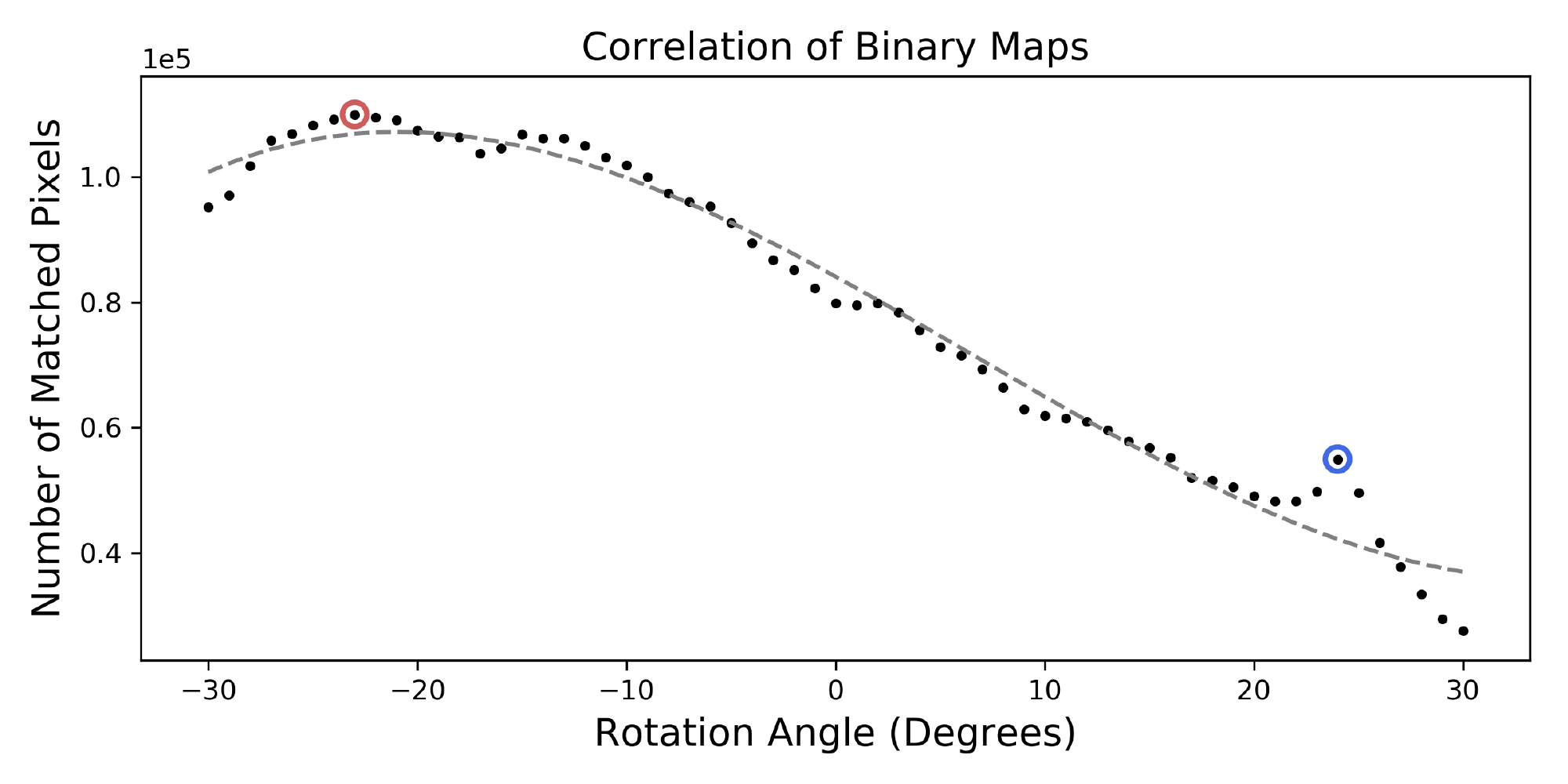}& \includegraphics[width=6cm]{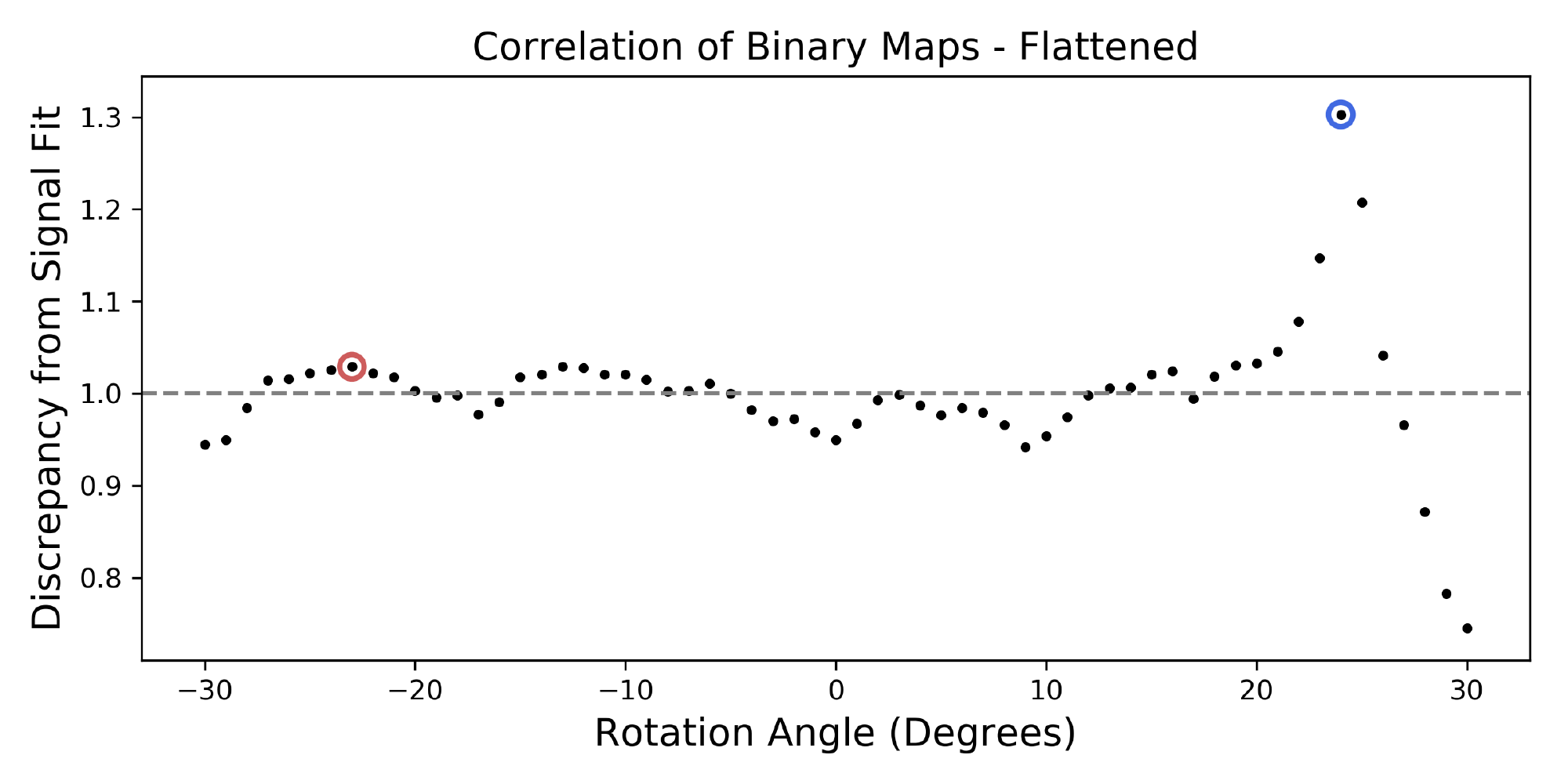} 
	\end{tabular}
	\caption{\red{Extra analysis needed to achieve a correct rotation degree match by locating local maxima. The blue circle indicates the correct rotation degree, the red circle indicates a possible false rotation degree match, and the grey dashed line represents the best fit. a) Results of the correlation between the two binary images for the Western hemisphere with a best fit line. The x-axis represents the rotation degree of the image and the y-axis corresponds to the number of pixels matched between the images at the degree. b) The correlation data for the Western hemisphere that has been smoothed by dividing the data by its best fit to bring out the local maxima at 24 degrees. The x-axis represents the rotation degree of the image and the y-axis relates to the discrepancy from the best fit with a value of 1 being an exact match.}}
	\label{fig:peaknorm}
\end{figure}

For the Eastern hemisphere of the Moon the rotation angle with the maximum number of pixels matched (highest correlation value) will be equivalent to the rotation angle of the final image. However, with the Western hemisphere, the maximum number of matched pixels is associated with an incorrect rotation angle due to Oceanus Procellarum acting as a false positive match. To combat this false match in rotation angle, AUGUR uses a third order polynomial fit to smooth the data and bring out local maxima that is associated with the correct rotation angle. This smoothed data and local maxima detection for the Western hemisphere correlation data is shown in Fig.~\ref{fig:peaknorm}.

After a rotation angle is obtained for the image, AUGUR can then complete the transformation of the coordinates from detector space (pixels) to selenographic coordinates ($\beta, \lambda$) using 

\begin{subequations}
	\begin{equation}
	\label{eq-a}
	\beta = sin^{-1} \left ( cos (c) sin (\beta_1) + \frac{y sin(c)cos(\beta_1)}{\rho}  \right )
	\end{equation}
for the latitude ($\beta$) and 
	
	\begin{equation}
	\label{eq-b}
	\lambda =\lambda_0+  tan^{-1} \left (\frac{x sin(c)}{\rho cos(\beta_1) cos(c)-y sin(\beta_1)sin(c)}  \right )
	\end{equation}
\end{subequations}
for the longitude ($\lambda$). Where $\rho = \sqrt{x^2 +y^2}$ and $c = sin^{-1}(\rho)$. Here ($x,y$) are the pixel coordinates of the LIF and ($\beta_1,\lambda_0$) are the sub-observer latitude and longitude previously obtained from \textit{JPL Horizons} \citep{Kopal1965}.
 
\section{Source of impactors}
\label{source}

The next step of the analysis is to identify the origin of the impactors, because it will provide information about their nature. In general, a meteoroid may originate from a \red{meteoroid} stream or from the background sporadic population of small near-Earth objects (NEOs). In the case of streams, the vast majority are of cometary origin and only a few are associated with asteroidal activity (e.g. Geminids). In order to estimate the kinetic energy ($KE$) of an impactor and subsequently its mass, it is essential to know the speed ($v_{im}$) at which it impacted onto the lunar surface and, if possible, the density of the object. The $v_{im}$ is obtained by linking the impactor to its source, either a known meteoroid stream or the sporadic background population. To link the meteoroid that produced a flash to a meteoroid stream, we use the methods described in \cite{avdellidou2019} and in \cite{madiedo2015}, after adjustments for the purposes of this work. 

The first thing is to check whether the impact happened during an outburst of a \red{meteoroid stream} or not. Then, we compute the solar longitude difference between the time of the impact flash and that of the \red{stream's} maximum activity. \red{In this work we do not use the solar longitudes that are provided in the literature but we calculated them using established methods \citep{steyaert1991}.} If such difference is above a selected threshold the stream is omitted. 
Next, the difference of solar longitudes is converted into a difference of time which is used to correct the \red{stream's} radiant equatorial coordinates (Ra, Dec)$_{ST}$ for the radiant drift. 
\red{After that, the vector going from the centre of the Moon to the impact point is converted from Moon-centred coordinates into ICRS coordinates (Ra, Dec)$_{imp}$ by taking into account the orientation of the Moon.} Then the angular distance between the corrected \red{stream} radiant coordinate (Ra, Dec)$_{ST}$ and (Ra, Dec)$_{imp}$ are computed, and is checked if such distance is below a selected threshold and if not, the \red{stream} is eliminated. If a \red{stream} passes all these stages is considered as a candidate.

Once a meteoroid stream is a candidate, we compute the speed at which a meteoroid from the said \red{stream} will impact the lunar surface. Using SPICE, we obtain the ephemerides of the Moon at the time of impact and compute the heliocentric distance of the Moon. This is used to determine the relative distance between the Moon and the \red{stream's} nodes (ascending, descending). By checking which distance is the closest, we retrieve the true anomaly of the closest node, which is used to compute the mean anomaly $M$ according to:
\begin{equation}\label{ana}
\begin{split}
 M=& \nu -2e\sin \nu +\left({\frac {3}{4}}e^{2}+{\frac {1}{8}}e^{4}\right)\sin 2\nu\\ 
 &-{\frac {1}{3}}e^{3}\sin 3\nu +{\frac {5}{32}}e^{4}\sin 4\nu +\cdots
\end{split}
\end{equation}
where $\nu$ is the true anomaly and $e$ the eccentricity of the \red{stream}. The reason we need the \red{stream's} mean anomaly is to construct its SPICE orbital elements array which consist of the \red{stream} relative information: perifocal distance, eccentricity, inclination, longitude of the ascending node, argument of periapse, mean anomaly at epoch, epoch (the instant at which the state of the body is specified by the elements) \red{and the Sun gravitational parameter (since orbits are heliocentric, J2000)}. All these orbital elements, except for the mean anomaly, are provided by the IAU Meteor Data Center\footnote{https://www.ta3.sk/IAUC22DB/MDC2007/}.
Using this array, we determine the epoch when the closest distance in the \red{stream's} orbit to the Moon is reached. This epoch is used to compute the heliocentric distance of the \red{stream's} meteoroids and compare it to that of the Moon in order to ensure that the difference is not bigger than 0.1~au. \red{This criterion value is based on orbital mechanics consideration: the difference in velocity vector makes sense only if the stream and the Moon are approximately at the same heliocentric distance \citep[see also][]{neslusan1998}.} Then we compute the difference between the velocity of the Moon at the epoch of the impact and that of the \red{stream's} meteoroids at the epoch of the closest distance. The resulted relative speed is the final $v_{im}$.

Next we eliminate the candidate streams that are geometrically impossible to produce the studied impact. For a given \red{stream} the lunar sub-radiant point is the point on the lunar surface where the radiant of the \red{stream} is directly above \citep{cudnik2009}. To compute the selenographic coordinates of the sub-radiant points, we first obtain the position of the Moon at the time of the impact in ECLIPJ2000. Next, using the \red{stream's} SPICE array (constructed earlier), we search its closest approach to the Moon, retrieve its coordinates in ECLIPJ2000 and determine the \red{stream's} position with respect to the centre of the Moon. The coordinates then are transformed to the Moon Principal Axes frame, which is a non-inertial body-fixed frame, where the z-axis is along the Moon's rotation axis. At this point, we have the \red{stream's} sub-radiant point ($x_{sub},y_{sub},z_{sub}$), and we compute the angular separation ($\theta$) from the LIF location. This is done by converting the planetocentric latitude and longitude of the LIF to rectangular coordinates ($x_f,y_f,z_f$) and compute the angle between the two vectors ($\theta$). Any meteoroid \red{stream} that has an angle between its sub-radiant point and an impact location larger than 90~deg is excluded. 

At this stage we probabilistically determine the most likely parent meteoroid stream, or whether it comes from the sporadic background population, following the procedure that is described in \cite{madiedo2015}. The probability of a stream being the source of a LIF is:
\begin{equation}
    p^{ST} = {{N^{ST}} \over {N^{ST} + N^{Other ST} + N^{SPO}}}
\end{equation}
$N^{ST}$ is the number of impacts per unit time that can be produced by a \red{stream}, $N^{Other ST}$ the number of impacts per unit time that can be produced by other active streams and $N^{SPO}$ is the number of impacts per unit time produced by the sporadic population. As for the probability of an impact flash coming \red{from} the sporadic population, it will be:
\begin{equation}
    p^{SPO} = 1 - \Sigma_i \;p^{ST_i}
\end{equation}
where $i$ represents the active streams that may be the source of the LIF. Following the assumptions of \cite{madiedo2015} (see paper for more details), the probability of a stream being the source of a LIF becomes:
\begin{equation}
    \label{prob}
    p^{ST} = {{\sigma \nu^{ST} \gamma^{ST}  cos(\theta) ZHR_{Earth}^{ST}(peak) 10^{-b|\lambda - \lambda_{peak}|}} \over {\nu^{SPO} \gamma^{SPO} HR_{Earth}^{SPO}+\sigma \nu^{ST} \gamma^{ST} cos(\theta) ZHR_{Earth}^{ST}(peak) 10^{-b|\lambda - \lambda_{peak}|} + \kappa}}
\end{equation}
with
\begin{equation}
    \kappa = \Sigma_i\; \sigma \nu^{ST}_i \gamma^{ST}_i cos(\theta_i) ZHR_{i,Earth}^{ST}(peak) 10^{-b_i|\lambda_i - \lambda_{i,peak}|}
\end{equation}
where 
\begin{itemize}
    \item $\sigma = {{d_{Earth}}\over {d_{Moon}}}$; $d$ is the distance to the radiant of the stream.
    \item $\nu = \left({{m_0 v_{im}^2}\over{2}}\right)^{s-1}E_{min}^{1-s}$; $v_{im}$ is the impact velocity, $m_0$ is the mass of a meteoroid \red{stream} producing on Earth a meteor of magnitude +6.5, $s$ is the mass index, which is related to the population index $r$ ($s = 1 + 2.5\,log\,r$) and $E_{min} = \eta\,E_{lum,min}$ is the minimum $KE$ detectable from the Earth with $\eta$ the luminous efficiency and $E_{lum,min}$ the minimum luminous energy detectable from the Earth. The selected values for $\eta$ are described immediately after, in $\S$\ref{masses}.
    \item $\gamma = {{\Phi_{Moon}}\over{\Phi_{Earth}}}$, with $\Phi = 1 +  \left({{v_{esc}}\over{V}}\right)^2$ as the gravitational focusing factor, $v_{esc}$ as the escape velocity of the central body and $v_{im}$ as the impact speed.
    \item $\theta$ is the angular distance between the sub-radiant point of the stream on the Moon and the location of the impact flash on the lunar surface.
    \item $ZHR_{Earth}^{ST}(peak)$ is the zenith hourly rate of the stream on Earth at its peak activity with $\lambda_{peak}$ to be the solar longitude at such activity peak.
    \item $\lambda$ is the solar longitude at the time of the LIF.
    \item $b$ is the slope governing the behaviour of $ZHR$ away from its peak \citep[][for more details]{jenniskens1994}.
    \item $HR_{Earth}^{SPO}$ is the average hourly rate of sporadic events.
\end{itemize}
In Eq.~\ref{prob}, the parameters $ZHR_{Earth}^{ST}(peak)$, $\lambda_{peak}$, $b$ and $r$ need to be specified for each potential stream where values are retrieved from the literature \citep{jenniskens1994,jenniskens2006,moorhead2019}, the International Meteor Organization (IMO)\footnote{https://www.imo.net/resources/calendar/} and the American Meteor Society (AMS)\footnote{https://www.amsmeteors.org/meteor-showers/2020-meteor-shower-list/}. 
Despite our extensive search to get these parameters for every stream we found to be a potential source for an impact flash, we still miss values for the above parameters of some minor streams. For those streams, we carried out  actions based on the missing parameter. In the case of the population index $r$, we examined all the available values for all the streams and selected the most probable one to be 2.5. As for the adopted $r$ value for the sporadic background, we selected it to be 3 \cite[see][and reference therein]{madiedo2015}. For $b$ values, we followed the convention set by \cite{jenniskens1994}. For minor streams that have high inclination (i $>$ 15~deg), which turned out to be all of them in our case, the value of $b$ can be set to 0.19. For $HR_{Earth}^{SPO}$ we assigned 10 meteors~$h^{-1}$  \citep[see][ and reference therein]{madiedo2015}. The $m_0$ is computed using the Eqs.~(1) and (2) from \cite{hughes1987}, $E_{lum,min}$ using the Eqs.~(5) \& (6) from \cite{suggs2014}, Eqs.~(18) from \cite{madiedo2015} and the equations described in $\S$\ref{masses_neliota}. The computation of $E_{lum,min}$ requires information about the systems used to observed each LIF such as the limiting magnitude and is obtained from the respective LIF publications. The results from the whole dataset, regarding the origin of an impactor and the respective probability are given in Tab.~\ref{table4}.

\section{Physical properties: mass and size}
\label{masses}
In this work we have collected LIF datasets from different projects (see $\S$\ref{data}) and we unite them in one single dataset in order to construct the size distribution. To do so, each dataset requires a different treatment.

During an impact event the $KE$ of the impactor is divided in several parts; one part is released as luminous energy, $E_{lum}$, and is detected by telescopes. This is used in order to derive the masses and later the sizes of the impacting meteoroids. The $E_{lum}$ is a small fraction $\eta$ of the $KE$ and thus the impacting mass is given by:
\begin{equation}
\label{ke}
m= \frac{2E_{lum}}{v_{im}^{2}\eta}  
\end{equation}
where the luminous efficiency $\eta$ has been estimated by several studies \citep{bellotrubio2000, moser2011, bouley2012}.

\subsection{NELIOTA}
\label{masses_neliota}

According to previous laboratory and observational studies, an impact flash is assumed to be a black body radiator and its spectral energy distribution is described by the Planck formula:
\begin{equation}
B(\lambda,T) = \frac{2hc}{\lambda^{5}}\frac{1}{\exp(\frac{hc}{\lambda k_{B}T}) - 1}
\label{plank}
\end{equation}
calculated in erg~cm$^{-2}$s$^{-1}$A$^{-1}$sr$^{-1}$, where $h=6.62\times10^{-27}$~g~cm$^{2}$s$^{-1}$ is the Planck constant, $c=3\times10^{10}$~cm~s$^{-1}$ is the speed of light, $k_{B}=1.38\times10^{-16}$~g~cm$^{2}$s$^{-2}$K$^{-1}$ is the Boltzmann constant, $T$ and $\lambda$ are the temperature of the flash and the wavelength of the photons, respectively. In addition, we accept that the observed flux $F$ for a given observing time where the Moon is at distance $D$ from the observer, comes from a half black body-like sphere of radius $r$ (in m) and thus is given by:

\begin{equation}
F= \frac{\pi r^{2}}{D^{2}} B(\lambda,T)
\end{equation}

\begin{figure}
	\centering
	\includegraphics[width=\linewidth]{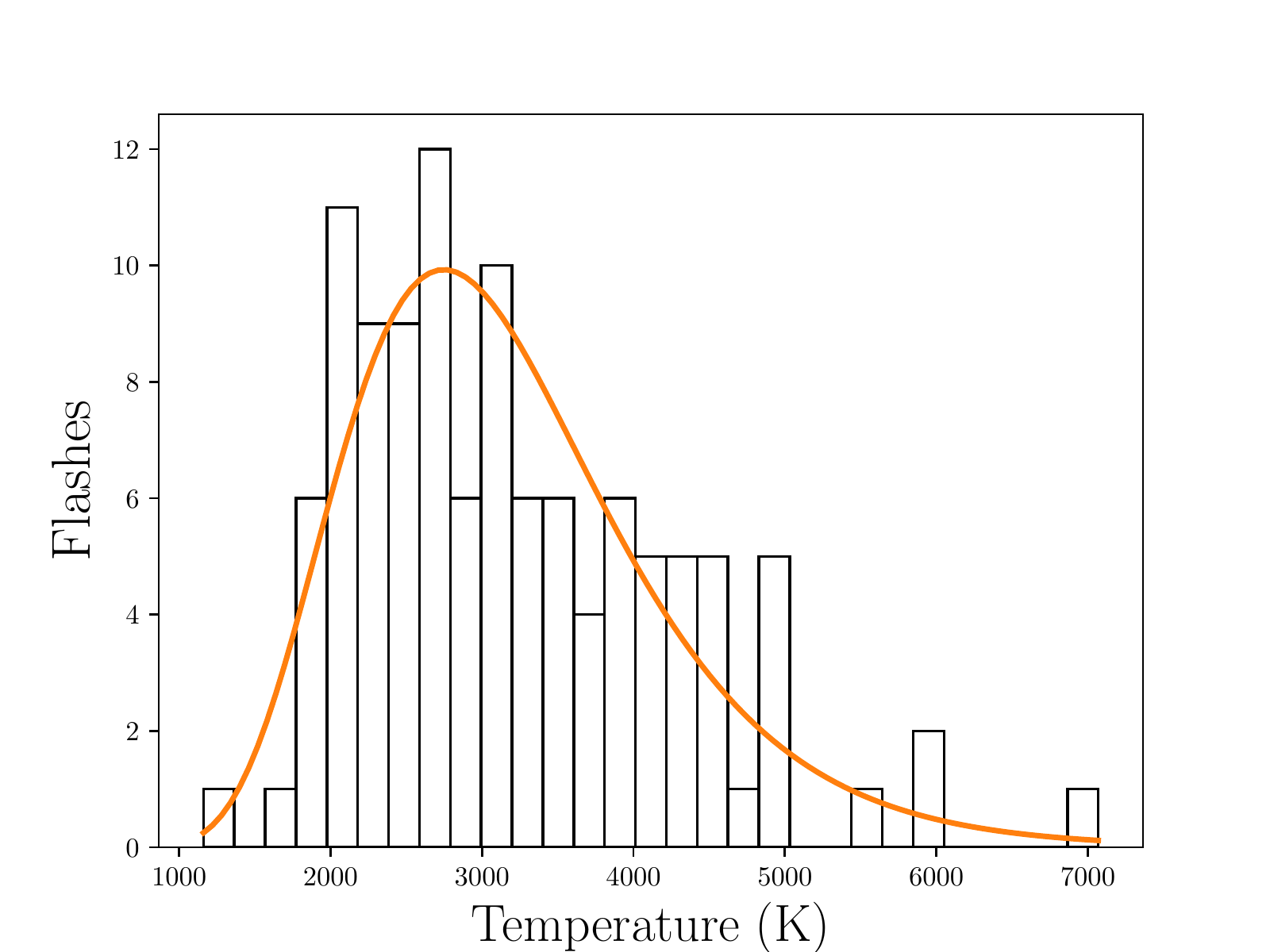}
	\caption{Temperature distribution of the observed flashes. The values correspond to the peak temperature measured from the initial $R$ and $I$ frames of each flash.}
	\label{temp}
\end{figure}

In this work we follow the same assumptions and methodology that are described in details by \cite{avdellidou2019} and here we remind the main steps. According to \cite{xilouris2018} the NELIOTA magnitudes $R$ and $I$ are calibrated in the Cousins system and the $R-I$ colour is described by the the following equations:

\begin{align}
\label{mags}
\begin{split}
R= -2.5 log (\frac{\pi r^2}{D^2}) -2.5 log(\xi_{R}) - 21.1 - ZP_{R} \\
I= -2.5 log (\frac{\pi r^2}{D^2}) -2.5 log(\xi_{I}) - 21.1 - ZP_{I}
\end{split}
\end{align}

\begin{equation}
\label{colour}
R-I= -2.5 log(\frac{\xi_{R}}{\xi_{I}}) - (ZP_{R} - ZP_{I})
\end{equation}
where $ZP_{R}$ = 0.555 and $ZP_{I}$ = 1.271, as defined by \cite{bessell1998}. The parameters $\xi$ are the spectral flux density of each band and are defined as:
\begin{equation}
\xi_R = \frac{\int_{}^{}B(\lambda,T) Rc(\lambda) d\lambda}{\int_{}^{}Rc(\lambda) d\lambda}~~~~~
\xi_I = \frac{\int_{}^{}B(\lambda,T) Ic(\lambda) d\lambda}{\int_{}^{}Ic(\lambda) d\lambda}
\end{equation}
where $Rc(\lambda)$ and $Ic(\lambda)$ are the responses of the Cousin filters.
The theoretical curve that relates the $R-I$ colour with flash temperature for this observational setup has been already estimated and presented in \cite{avdellidou2019}. For each flash, given the $R-I$ we obtain its $T$ and we can evaluate the $\xi_R$ and $\xi_I$. Evaluating Eqs.~\ref{mags}, we compute for each event the $T$ of the LIF (Fig.~\ref{temp}) and the radius $r$ of the theoretical area of emission. 
To estimate the error in $T$, a Monte Carlo approach is adopted with typically $10^5$ iterations. At each iteration, a temperature value is derived from a new set of $R$ and $I$ that are extracted from Gaussian distributions centred on the nominal $R$ and $I$ values of the LIF and standard deviations equal to the magnitude errors given by NELIOTA and/or literature \citep{liakos2020}. It has been stated that the NELIOTA $R$ and $I$ magnitude values and errors, as presented in the web-page, are an approximation and therefore when possible we use the published ones \citep{xilouris2018, liakos2020}.

Knowing now the temperatures and radii of the luminous events we can calculate the $E_{lum}$ from:
\begin{equation}
\label{elum}
E_{lum} = Lt
\end{equation}
where $t$ the duration of the luminous event, $L$ the luminosity that is given by:
\begin{equation}
\label{l}
L = \pi^{2}r^{2}\int_{400}^{900}B(\lambda, T)d\lambda
\end{equation}
and finally we evaluate Eq.\ref{ke}. 
For the impact flashes that are detected in both cameras in at least two frames we derive the $R$ and $I$ magnitudes (see Tab.~\ref{table1}) for all frames using the Source Extractor tool \citep{1996A&AS..117..393B} as described in detail in \cite{avdellidou2019}. For each of those frames the $E_{lum}$ is calculated and summed in order to evaluate the Eqs.~\ref{elum}\&\ref{l}. Table~\ref{table3} contains the derived $T$ and $E_{lum}$ for the NELIOTA data.

For the impacts that are associated with \red{meteoroid} streams we use the $v_{im}$ as estimated before, whereas for impacts that come from the background population we assume a velocity of 24~km~s$^{-1}$ \citep{mcnamara2004}. The $\eta$ values were selected to be $\eta_1$=$1.5\times10^{-3}$ and $\eta_2$=$5\times10^{-4}$ \citep{bouley2012} and thus two mass datasets are produced (Tab.~\ref{table4}). Here we have to clarify that the same $\eta$ values were used to evaluate Eq.~\ref{prob} and make the link to either the background population or a meteor \red{stream}. 

\begin{table}
\centering
\caption{Multi-frame events. Magnitudes of the initial frames were taken from the literature or NELIOTA website, while the rest were calculated as described in \protect\cite{avdellidou2019}. $T$ is estimated as described in $\S$\ref{masses_neliota}.}
\label{table1}
\begin{tabular}{r|ccc}
\hline
\hline
 Flash ID/   &   $R \pm \sigma_R$ 	& $I \pm \sigma_I$  & $T \pm \sigma_T$ \\  
 frame & (mag) 			& 	(mag) 			&  (K)\\
 \hline
2/a	&	6.67	$\pm$	0.07	&	6.07	$\pm$	0.06	&	4429	$\pm$	326	\\
2/b	&	10.01$\pm$	0.17	&	8.26	$\pm$	0.07	&	1941	$\pm$	203	\\
13/a	&	8.27	$\pm$	0.04	&	6.32	$\pm$	0.01	&	1803	$\pm$	14	\\
13/b	&	9.43	$\pm$	0.12	&	7.44	$\pm$	0.02	&	1736	$\pm$	87	\\
19/a	&	9.17	$\pm$	0.07	&	8.07	$\pm$	0.03	&	2991	$\pm$	149	\\
19/b	&	13.00$\pm$	1.50	&	8.96	$\pm$	0.08	&	2228	$\pm$	1141	\\
20/a	&	8.52	$\pm$	0.03	&	7.04	$\pm$	0.01	&	2261	$\pm$	44	\\
20/b	&	10.01$\pm$	0.14	&	8.27	$\pm$	0.07	&	1944	$\pm$	176	\\
20/c	&	11.32$\pm$	0.25	&	9.14	$\pm$	0.08	&	1634	$\pm$	174	\\
40/a	&	9.16	$\pm$	0.09	&	7.73	$\pm$	0.02	&	2333	$\pm$	139	\\
40/b	&	11.19$\pm$	0.53	&	9.68	$\pm$	0.11	&	2433	$\pm$	963	\\
42/a	&	8.78	$\pm$	0.05	&	7.74	$\pm$	0.02	&	3019	$\pm$	108	\\
42/b	&	11.78$\pm$	1.52	&	8.85	$\pm$	0.05	&	2583	$\pm$	1492	\\
47/a	&	8.36	$\pm$	0.04	&	7.30	$\pm$	0.02	&	3089	$\pm$	116	\\
47/b	&	11.02$\pm$	0.30	&	8.81	$\pm$	0.03	&	1667	$\pm$	194	\\
51/a	&	8.50	$\pm$	0.11	&	7.16	$\pm$	0.02	&	2653	$\pm$	134	\\
51/b	&	7.84	$\pm$	0.07	&	6.60	$\pm$	0.02	&	2654	$\pm$	133	\\
51/c	&	8.98	$\pm$	0.14	&	7.45	$\pm$	0.03	&	2205	$\pm$	197	\\
51/d	&	10.02$\pm$	0.25	&	8.21	$\pm$	0.04	&	1907	$\pm$	250	\\
62/a	&	10.12$\pm$	0.20	&	9.29	$\pm$	0.10	&	3721	$\pm$	724	\\
62/b	&	10.9	$\pm$	0.30	&	9.02	$\pm$	0.10	&	1862	$\pm$	298	\\
65/a	&	6.65	$\pm$	0.10	&	5.49	$\pm$	0.06	&	2784	$\pm$	241	\\
65/b	&	8.06	$\pm$	0.23	&	6.68	$\pm$	0.07	&	2424	$\pm$	378	\\
69/a	&	9.64	$\pm$	0.16	&	8.21	$\pm$	0.07	&	2312	$\pm$	250	\\
69/b	&	10.13$\pm$	0.30	&	8.21	$\pm$	0.70	&	2363	$\pm$	928	\\
71/a	&	8.95	$\pm$	0.13	&	8.02	$\pm$	0.07	&	3371	$\pm$	438	\\
71/b	&	10.94$\pm$	0.40	&	9.11	$\pm$	0.09	&	1964	$\pm$	426	\\
94/a	&	8.40	$\pm$	0.10	&	5.6	$\pm$	0.01	&	1161	$\pm$	40	\\
94/b	&	12.7	$\pm$	1.10	&	8.34	$\pm$	0.04	&	843	$\pm$	335	\\
\hline
\hline
\end{tabular}
\end{table}

\subsection{Marshall Space Flight Center}
\label{masses_NASA}
The analysis of the data from the Marshall Space Flight Center is simpler. As described before, the input data for this work are the LIFs presented in \cite{suggs2014}. We evaluate Eq.~\ref{ke} using the peak $E_{lum}$ that is presented in their Table~1 (and we repeat in Tab.~\ref{table3}) and the $v_{im}$ we derived by re-estimating the origin (meteor stream or background) of each of their flashes (see $\S$\ref{source}). The final masses for this dataset were derived by using the $\eta_S$ expression as described in \citep{suggs2014} and given by the Eq.~(8) of that study. In addition, masses and sizes were derived using the $\eta_1$ and  $\eta_2$ values for comparison purposes. The same $\eta$ were also used to evaluate Eq.~\ref{prob} for Marshall Space Flight Center dataset (Tab.~\ref{table4}).

\subsection{Other observations}
\label{masses_others}
For the rest of the datasets presented in $\S$\ref{data} we follow two paths. For the events that have published $E_{lum}$ we use directly these values \citep{larbi2015}, while for the rest we estimate the $E_{lum}$ using the method described in \cite{madiedo2015}. The $E_{lum}$ values that we adopted or derived are presented in Tab.~\ref{table3}. Again we use $\eta_1$ and $\eta_2$ values, while the $v_{im}$ for each flash is derived again in $\S$\ref{source} with the same method as for the other datasets (Tab.~\ref{table4}). 

\subsection{Size frequency distribution of lunar impactors.}
\label{sizes}

\begin{figure}
	\centering
	\includegraphics[width=\linewidth]{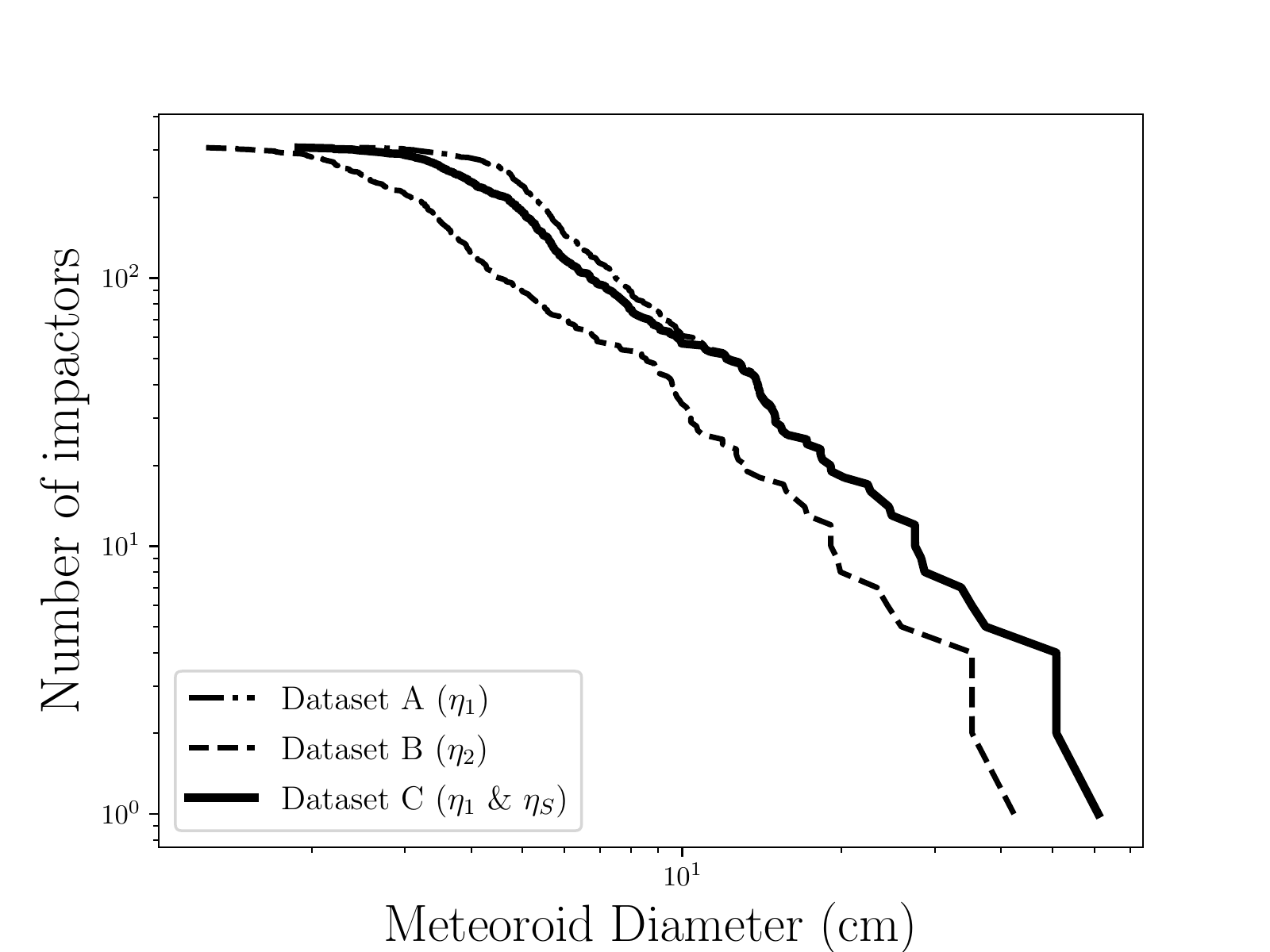}
	\caption{Dataset A contains meteoroid diameters form all detection that were analysed using $\eta_1=5\times10^{-4}$ and Dataset B diameters that were analysed using $\eta_2=1.5\times10^{-3}$. Dataset C is made of diameters that were produced using $\eta_1$ for NELIOTA and other minor observations, while data from \cite{suggs2014} \red{were} analysed using their own $\eta_S$ estimation.}
	\label{sfd}
\end{figure}

The masses are converted to sizes assuming spherical-shaped meteoroids and using density estimations from the literature \citep{babadzhanov2009}. For the events that are linked to eta Aquarids (ETA) we assign the same density that is estimated for Orionids $\delta$=0.9~gr~cm$^{-3}$ since both streams originate from comet Halley. For the meteoroids that originate from the background population or from sources that have no density estimations we assume $\delta$=1.8~gr~cm$^{-3}$ \citep{avdellidou2019}.
Finally we group the diameters of all the 307 impactors (see Tab.~\ref{table4}) and construct a global size frequency distribution (SFD) (Fig.~\ref{sfd}).

\section{First LIF detection from Observatoire de la C\^ote d'Azur, Nice, France}
\label{flash_oca}

\begin{figure}
	\centering
	\includegraphics[width=\linewidth]{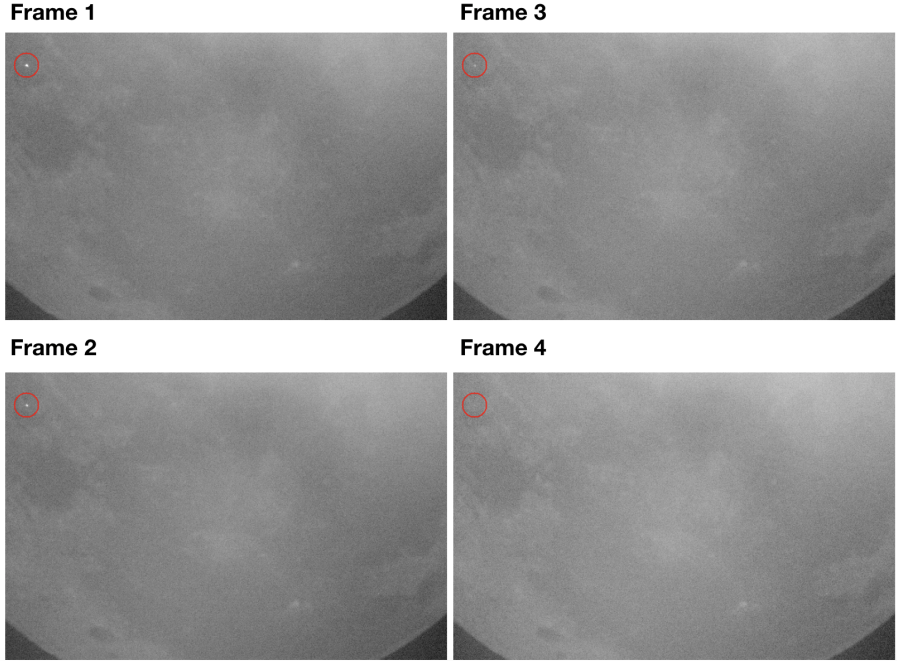}
	\caption{The evolution of the first lunar impact flash detected from France. The detection took place on the night of May 27$^{th}$, 2020 at 20:48:49.420 UTC from OCA, Mt. Gros site.}
	\label{flash}
\end{figure}

In order to test the aforementioned detection and analysis methods we performed observations during May 2020. The telescope used was a MEADE ACF 40~cm equipped with a CMOS ASI ZWO 183 mono camera. The telescope was guiding on the lunar crescent using the lunar autoguider that our team developed specifically for this project.
The frame rate was 20 fps and the integration time was 0.05 seconds. On the night of May 27$^{th}$, 2020 at 20:48:49.420 UTC, we detected our first impact flash (see Fig.~\ref{flash}) from the Observatoire de la Côte d’Azur and the site of Mt. Gros in Nice. This is the first live impact observed for the project ``Flash!", the first from the Observatoire de la Côte d’Azur and the first in France.
Our algorithm that is described in $\S$\ref{detection} detected a luminous event that lasted for four frames and thus has a duration of at least 0.2 seconds. According to our method described in $\S$\ref{coords} we estimate the impact coordinates to be: $\beta$=-36.3$\pm$1 and $\lambda$=-25.4$\pm$1.
Although during the testing phase we performed observations only with one camera, we are confident that the luminous event that we detected is due to a LIF. Supporting factors is the long duration and the constant coordinates in all the frames. Following the procedure described in $\S$\ref{source} we conclude that the meteoroid originated from the sporadic population.

\section{Discussion}
\label{discussion}

Comparing the new temperature distribution that we derive in this work, using 112 NELIOTA LIFs, with the previous one of \cite{avdellidou2019} we see that the maximum $T$ derived ($\sim7,000$~K) is significantly higher. The updated data follow a log-normal distribution with mode $\sim$2,750~K. The updated mode is slightly higher than the 2,550~K that was presented in \cite{avdellidou2019} and essentially the same with the theoretical estimation of 2,800~K \citep{nemtchinov1998}. However, this difference with the previous analysis may not be statistically significant. The vast majority of the flashes appear to have a $T<5,000$~K and only four events exceed this value. 
However, the discrepancy between the theoretical maximum $T$ \cite[3,776~K,][]{cintala1992} and the observed LIF $T$ still remains. According to early laboratory experiments \citep{eichhorn1975,eichhorn1976} the impactors density may affect the flash temperatures but from the current dataset we do not observe a preference of impactors from a specific stream (and thus a specific composition and density) to produce those high $T$. 

Regarding the mass estimation of the impactors there is a number of reasons that may affect the estimated values and subsequently the final slope of the SFD.
First of all the measured $E_{lum}$ may be underestimated. For example, for some frames luminosity can be below the background so it cannot be detected. Another reason, applied to NELIOTA data, is due to the fact that we take into account only the frames where the flash is detected in both cameras. All LIFs - even the single-frame ones - are detected in at least one more frame, usually in the $I$ filter. Unfortunately, since the detection is not confirmed in the respective $R$ filter frame we do not use it for the $E_{lum}$ calculation, making the total $E_{lum}$ of the flashes to be definitely underestimated. One could imagine that if we know the cooling rates we can correct the energy. However, the cooling rate is not the same for every LIF \citep{avdellidou2019}, meaning that the $E_{lum}$ ratio between the frames in the multiframe events is not constant. This prevents us from making assumptions for the lost energy.

The most important factor that can change the mass estimation, even by an order of magnitude, is the luminous efficiency $\eta$. What $\eta$ may also affect is the link to stream (or not) as it is a factor on which Eq.~\ref{prob} depends. To understand if this dependence is significant we compare the probability results from both calculations. As it is seen in Tab.~\ref{table2} by changing the luminous efficiency the link changes only for 5 LIFs out of 112 which we recon is not significant and thus may not affect the final SFD. 

\begin{table}
\centering
\caption{The change in impactor's origin due to the different assumption for the luminous efficiency $\eta$ (ID 6, 11, 18, 73 and 74) and due to differences in selenographic coordinate estimation (ID 17 and 18). NC is the nominal NELIOTA coordinates as described in $\S$\ref{data}, UC the updated ones as derived in this work, $\eta1=5\times10^{-4}$ and $\eta2=1.5\times10^{-3}$. The results for the total dataset are presented in Tab.~\ref{app_table2}.}
\label{table2}
\begin{tabular}{|r|cccc|}
\hline
\hline
 ID &		NC 	&	NC  &		UC 	&	UC\\
		& $\eta_1$ & $\eta_2$ & $\eta_1$ &  $\eta_2$\\
\hline
6	&	ETA		&	SPO		&	ETA		&	SPO\\
11	&	PER		&	SDA		&	PER		&	SDA\\
17	&	SPO		&	SPO		&	NIA		&	SPO\\
18	&	NIA		&	SPO		&	SPO		&	SPO\\
73	&	CAP		&	SPO		&	CAP		&	SPO\\
74	&	CAP		&	SPO		&	CAP		&	SPO\\
\hline
\hline
\end{tabular}
\end{table}

The link of an impactor to a specific meteoroid stream or to the sporadic population affects also the mass estimation as the speed alters significantly and may vary between 17 and 70~km/s! \red{In this work we advanced from \cite{suggs2014} by checking the full database of meteoroid streams and not only the strongest ones and we also advanced from \cite{avdellidou2019} as we investigate also the probability of the impactor to originate from the background population.} Since we have two datasets of selenographic coordinates (Tab.~\ref{app_table1}) one obtained from the literature and the other produced in this work, at least for the NELIOTA data, we compute the probabilities for both datasets (Tab.~\ref{app_table2}). In principle any change in selenographic coordinates may place a LIF inside or outside of the lunar area on which a given meteoroid \red{stream} can impact. As it is seen in Tab.~\ref{table2}, by changing the selection of the selenographic coordinates the link changes only for two out of 112 LIFs. This is an insignificant fraction that will not affect the global SFD of lunar impactors. \red{The source probability for each impactor may be affected by the selection of the parameters $b$, $r$ and $m$ (for Eq.~\ref{prob}) that characterise each stream and are not always precisely known for each one. The most important uncertainty is introduced by the impact speed which is required in order to derive the meteoroid's mass. In the case where an impactor is linked to a meteoroid stream this problem is lifted. 
However, in the most numerous cases where the impactors are linked to the background population an average speed value is selected to be applied to all the events. Several average values have been adopted so far from the different LIF studies, ranging from 17 to 24~km~s$^{-1}$ \citep{ortiz1999, ortiz2000, ortiz2006, suggs2014, madiedo2014}. According to the impact speed distribution on the lunar surface by \cite{lefeuvre2011} a minimum value can be as low as $\sim$5~km~s$^{-1}$ while the tail of the distribution can reach $\sim$40~km~s$^{-1}$. The distribution of the impacting speeds derived by \cite{mcnamara2004} agrees with the low limit of \cite{lefeuvre2011} but at the high end there is a longer tail where speeds can reach up to 65~km~s$^{-1}$, with the average value being $\sim$24~km~s$^{-1}$. Therefore, in reality a meteoroid's speed can deviate from this average, resulting in masses up to 25 times larger or down to nine times smaller than the ones presented in tis work, affecting the SFD.}

Since the selection of $\eta$ affects the final SFD we need to be careful on the way we link the data from the different sources. As it is seen in Fig.~\ref{sfd} the SFD slope in the very small impactor sizes is slightly affected if we do not use the $\eta$ estimation of \cite{suggs2014}. Therefore, in order to be consistent with their study and results we adopt the Dataset C which uses the Marshall Space Flight Centre data processed with the $\eta$ estimation for their own observing system, while for the rest we use the $\eta$ values taken from the literature. 

\red{\cite{halliday1996}, using the fireball data from the Canadian camera network, derived an SFD of small impactors, breaking around the size of 10-15~cm in two segments with different slopes, those being $\sim$-1.3 for the smaller and $\sim$-3 for the larger objects. The SFD (Dataset 3 of Fig.~\ref{sfd}) that we derive here appears also to have a small knee at the same sizes and corresponding slopes -1.4 and -2.2 respectively, while the vast majority of the meteoroids (80$\%$) appear to have sizes $<$10~cm.}

\section{Conclusions}
In this work we present all the computational steps to (i) detect the lunar flashes that correspond to meteoroid impact events; (ii) identify their selenographic impact coordinates, that will later serve in the detection of the impact craters; (iii) to link an impactor to a meteoroid stream or to the background sporadic population; (iv) estimate their masses and sizes. We tested the previous steps to a suite of LIF events presented in the literature in the last 20 years. By unifying different datasets from various observers, requiring different approach in their analysis due to different observing methods, we constructed the lunar impactors' SFD of small sizes (cm-dm). Furthermore we applied the steps above in our own observations during the testing phase of our observing project with a result the detection of the first LIF for the Observatoire de la C\^ote d'Azur.
\section*{Acknowledgements}
This work was supported by the Programme National de Plan\'etologie (PNP) of CNRS/INSU, co-funded by CNES and by the program ``Flash!" supported by Cr\'edits Scientifiques Incitatifs (CSI) of the Universit\'e Nice Sophia Antipolis. CA was supported by the French National Research Agency under the project ``Investissements d'Avenir" UCA$^\text{JEDI}$ with the reference number ANR-15-IDEX-01. Part of this work was also supported by the Planetary Astronomy program of the National Aeronautics and Space Administration, as well as NASA’s Lunar Reconnaissance Orbiter project. We would like to thank E. Bondoux for helping to access the telescope facility at Observatoire de la C\^ote d'Azur.

\newpage
\small
\begin{longtable}{|r|ccr||r|ccr|}
\caption{\label{table3} The total dataset that is used in this work. References are: Flash ID 1-112 from NELIOTA website; 182-307 from \cite{suggs2014}; 113, 114, 116, 120 from \cite{dunham2000};115, 118, 119, 120, 122 from \cite{ortiz2000}; 126-129 from \cite{ortiz2002}; 132-134 from \cite{ortiz2006}; 136-139, 144, 171-177 from \cite{ortiz2015}; 117, 130, 131 from \cite{cudnik2003}; 123-125 from \cite{yanagisawa2002}; 135 from \cite{yanagisawa2006}; 151, 159-170 from \cite{madiedo2015}; 140-143, 145-150 from \cite{madiedo2015b}; 154-158 from \cite{madiedo2016}; 180 from \cite{madiedo2018}; 152, 153, 178, 179, 181 from \cite{larbi2015}. } \\
\hline\hline
 ID	&	Date-Time (UT)	&	T (K)	&	E$_{lum}$ (J) &  ID	&	Date-Time (UT)	&	T (K)	&	E$_{lum}$ (J)	\\
\hline
\endfirsthead
\multicolumn{8}{c}
        {{ Table \thetable\ : Continued from previous page.}} \\
\hline\hline
\endhead
\hline\endfoot
1	&	20170201-17:13:57.863	&	2969 $\pm$ 276	&	14743	&	155	&	20130417-21:38:02.000	&		&	254186	\\
2	&	20170301-17:08:46.573	&	4429 $\pm$ 326	&	334934	&	156	&	20130417-22:12:25.000	&		&	254186	\\
3	&	20170301-17:13:17.360	&	3446 $\pm$ 367	&	34803	&	157	&	20130417	-22:39:00.000	&		&	87927	\\
4	&	20170304-20:51:31.853	&	4168 $\pm$ 714	&	27887	&	158	&	20130418-22:25:50.000	&		&	363078	\\
5	&	20170401-19:45:51.650	&	2148 $\pm$ 178	&	17069	&	159	&	20130811-20:16:33.000	&		&	405394	\\
6	&	20170501-20:30:58.137	&	2474 $\pm$ 310	&	17844	&	160	&	20130811-20:43:19.000	&		&	80695	\\
7	&	20170627-18:58:26.680	&	1929 $\pm$ 340	&	17107	&	161	&	20130811-21:08:14.000	&		&	534413	\\
8	&	20170628-18:45:25.803	&	3134 $\pm$ 1050	&	41286	&	162	&	20130812-19:49:57.000	&		&	1580881	\\
\hline
\hline
\end{longtable}

\newpage
\begin{landscape}
\begin{longtable}{|r|ccccc|ccccc|ccccc|}
\caption{\label{table4}Summary of the characteristics of the impactors. The meteoroid source (S) with its probability (P), its impact speed ($v_{imp}$) \red{in km~s$^{-1}$}, mass ($m$) \red{in gr} and size ($d$) \red{in cm} are calculated using three different estimations for the luminous efficiency $\eta$. We use $\eta_1=5\times10^{-4}$ (S$_1$, P$_1$, $v_{imp1}$, $m_1$, $d_1$) and $\eta_2=1.5\times10^{-3}$ (S$_2$, P$_2$, $v_{imp2}$, $m_2$, $d_2$) for the whole dataset. Additionally, for the data taken from \cite{suggs2014} we use the expression of $\eta_S$ that is described in their work (S$_S$, P$_S$, $v_{impS}$, $m_S$, $d_S$). }\\
\hline\hline
ID & S$_1$	&	P$_1$	&	$v_{imp1}$	&	$m_1$	&	$d_1$	&	S$_2$	&	P$_2$	&	$v_{imp2}$	&	$m_2$	&	$d_2$	&	S$_S$	&	P$_S$	&	$v_{impS}$	&	$m_S$ 	&	$d_S$	\\
\hline
\endfirsthead
\multicolumn{16}{c}
        {{ Table \thetable\ : Continued from previous page.}} \\
\hline\hline
\endhead
\hline\endfoot
1	&	SPO	&	0.79	&	24.00		&	102.38	&	4.77	&	SPO	&	0.82	&	24.00		&	34.13	&	3.31	&		&		&		&		&		\\
2	&	SPO	&	0.99	&	24.00		&	2325.93	&	13.51&	SPO	&	1.00	&	24.00		&	775.31	&	9.37	&		&		&		&		&		\\
3	&	SPO	&	1.00	&	24.00		&	241.69	&	6.35	&	SPO	&	1.00	&	24.00		&	80.56	&	4.41	&		&		&		&		&		\\
4	&	SPO	&	0.99	&	24.00		&	193.66	&	5.90	&	SPO	&	0.99	&	24.00		&	64.55	&	4.09	&		&		&		&		&		\\
5	&	SPO	&	1.00	&	24.00		&	118.53	&	5.01	&	SPO	&	1.00	&	24.00		&	39.51	&	3.47	&		&		&		&		&		\\
6	&	ETA	&	0.53	&	64.45		&	17.18	&	3.32	&	SPO	&	0.54	&	24.00		&	41.31	&	4.44	&		&		&		&		&		\\
7	&	SPO	&	0.95	&	24.00		&	118.80	&	5.01	&	SPO	&	0.96	&	24.00		&	39.60	&	3.48	&		&		&		&		&		\\
8	&	SPO	&	0.87	&	24.00		&	286.71	&	6.73	&	SPO	&	0.90	&	24.00		&	95.57	&	4.66	&		&		&		&		&		\\
\hline
\hline
\end{longtable}
\end{landscape}

\newpage
\appendix

\section{Additional data tables}

\begin{longtable}{|r|cccc||r|cccc|}
\caption{\label{app_table1}NC are the nominal NELIOTA coordinates as described in $\S$\ref{data} and UC the updated ones as derived in this work.}\\
\hline
\hline
 ID &		Long.NC 	&	Lat.NC  &	Long.UC 	&	Lat.UC &  ID &	Long.NC 	&	Lat.NC  &	Long.UC 	&	Lat.UC\\
\hline
\endfirsthead
\multicolumn{10}{c}
        {{ Table \thetable\ : Continued from previous page.}} \\
\hline\hline
\endhead
\hline\endfoot
1	&	-1.5	&	-29.2	&	-1.01		&	-29.09	&	29	&	13.0	&	50.0	&	10.49	&	52.28	\\
2	&	-10.3	&	-9.7	&	-8.93		&	-9.70		&	30	&	-36.9	&	73.4	&	-42.58	&	73.28	\\
3	&	4.5	&	-29.9	&	-12.79	&	-29.62	&	31	&	-40.7	&	79.2	&	-44.93	&	78.88	\\
4	&	-12.7	&	-58.9	&	-10.58	&	-59.611	&	32	&	-1.4	&	-52.0	&	0.99	&	-49.80	\\
5	&	11.6	&	-58.8	&	13.92	&	-58.26	&	33	&	21.7	&	74.5	&	22.81	&	74.775	\\
6	&	4.7	&	-43.2	&	5.978	&	-44.89	&	34	&	4.3	&	24.6	&	6.92	&	25.47	\\
7	&	26.8	&	-22.5	&	26.99	&	-22.96	&	35	&	3.6	&	-59.0	&	2.18	&	-59.60	\\
8	&	5.6	&	0.0	&	11.09	&	-1.25		&	36	&	17.4	&	-58.2	&	18.96	&	-56.42	\\
\hline
\hline
\end{longtable}

\newpage
\begin{longtable}{|r|cccc||r|cccc|}
\caption{\label{app_table2} Origin for NELIOTA impactors are estimated by using the nominal NELIOTA coordinates (NC) as described in $\S$\ref{data} and the updated ones as derived in this work (UC), for $\eta1=5\times10^{-4}$ and $\eta2=1.5\times10^{-3}$.}\\
\hline
\hline
 ID &		NC 	&	NC  &	UC 	&	UC &  ID &	NC 	&	NC  &	UC 	&	UC\\
	& $\eta_1$ & $\eta_2$ & $\eta_1$ &  $\eta_2$ & 	&$\eta_1$ & $\eta_2$ & $\eta_1$ &  $\eta_2$\\
\hline
\endfirsthead
\multicolumn{10}{c}
        {{ Table \thetable\ : Continued from previous page.}} \\
\hline\hline
\endhead
\hline\endfoot
1	&	SPO	&	SPO	&	SPO	&	SPO	&	29	&	GEM	&	GEM	&	GEM	&	GEM	\\
2	&	SPO	&	SPO	&	SPO	&	SPO	&	30	&	SPO	&	SPO	&	SPO	&	SPO	\\
3	&	SPO	&	SPO	&	SPO	&	SPO	&	31	&	SPO	&	SPO	&	SPO	&	SPO	\\
4	&	SPO	&	SPO	&	SPO	&	SPO	&	32	&	SPO	&	SPO	&	SPO	&	SPO	\\
5	&	SPO	&	SPO	&	SPO	&	SPO	&	33	&	SPO	&	SPO	&	SPO	&	SPO	\\
6	&	ETA	&	SPO	&	ETA	&	SPO	&	34	&	SPO	&	SPO	&	SPO	&	SPO	\\
7	&	SPO	&	SPO	&	SPO	&	SPO	&	35	&	SPO	&	SPO	&	SPO	&	SPO	\\
8	&	SPO	&	SPO	&	SPO	&	SPO	&	36	&	SPO	&	SPO	&	SPO	&	SPO	\\
\hline
\hline
\end{longtable}


\bibliographystyle{model5-names}\biboptions{authoryear}
\bibliography{references,mypapers}

\begin{thebibliography}{79}
\expandafter\ifx\csname natexlab\endcsname\relax\def\natexlab#1{#1}\fi
\providecommand{\url}[1]{\texttt{#1}}
\providecommand{\href}[2]{#2}
\providecommand{\path}[1]{#1}
\providecommand{\DOIprefix}{doi:}
\providecommand{\ArXivprefix}{arXiv:}
\providecommand{\URLprefix}{URL: }
\providecommand{\Pubmedprefix}{pmid:}
\providecommand{\doi}[1]{\href{http://dx.doi.org/#1}{\path{#1}}}
\providecommand{\Pubmed}[1]{\href{pmid:#1}{\path{#1}}}
\providecommand{\bibinfo}[2]{#2}
\ifx\xfnm\relax \def\xfnm[#1]{\unskip,\space#1}\fi
\bibitem[{{Ait Moulay Larbi} et~al.(2015){Ait Moulay Larbi}, {Daassou},
  {Baratoux}, {Bouley}, {Benkhaldoun}, {Lazrek}, {Garcia} \&
  {Colas}}]{larbi2015}
\bibinfo{author}{{Ait Moulay Larbi}, M.}, \bibinfo{author}{{Daassou}, A.},
  \bibinfo{author}{{Baratoux}, D.}, \bibinfo{author}{{Bouley}, S.},
  \bibinfo{author}{{Benkhaldoun}, Z.}, \bibinfo{author}{{Lazrek}, M.},
  \bibinfo{author}{{Garcia}, R.}, \& \bibinfo{author}{{Colas}, F.}
  (\bibinfo{year}{2015}).
\newblock \bibinfo{title}{{First Lunar Flashes Observed from Morocco (ILIAD
  Network): Implications for Lunar Seismology}}.
\newblock {\it \bibinfo{journal}{Earth Moon and Planets}\/},  {\it
  \bibinfo{volume}{115}\/}, \bibinfo{pages}{1--21}.
  \DOIprefix\doi{10.1007/s11038-015-9462-1}.
\bibitem[{{Arakawa} et~al.(2020){Arakawa}, {Saiki}, {Wada}, {Ogawa}, {Kadono},
  {Shirai}, {Sawada}, {Ishibashi}, {Honda}, {Sakatani}, {Iijima}, {Okamoto},
  {Yano}, {Takagi}, {Hayakawa}, {Michel}, {Jutzi}, {Shimaki}, {Kimura},
  {Mimasu}, {Toda}, {Imamura}, {Nakazawa}, {Hayakawa}, {Sugita}, {Morota},
  {Kameda}, {Tatsumi}, {Cho}, {Yoshioka}, {Yokota}, {Matsuoka}, {Yamada},
  {Kouyama}, {Honda}, {Tsuda}, {Watanabe}, {Yoshikawa}, {Tanaka}, {Terui},
  {Kikuchi}, {Yamaguchi}, {Ogawa}, {Ono}, {Yoshikawa}, {Takahashi}, {Takei},
  {Fujii}, {Takeuchi}, {Yamamoto}, {Okada}, {Hirose}, {Hosoda}, {Mori},
  {Shimada}, {Soldini}, {Tsukizaki}, {Iwata}, {Ozaki}, {Abe}, {Namiki},
  {Kitazato}, {Tachibana}, {Ikeda}, {Hirata}, {Hirata}, {Noguchi} \&
  {Miura}}]{arakawa2020}
\bibinfo{author}{{Arakawa}, M.}, \bibinfo{author}{{Saiki}, T.},
  \bibinfo{author}{{Wada}, K.}, \bibinfo{author}{{Ogawa}, K.},
  \bibinfo{author}{{Kadono}, T.}, \bibinfo{author}{{Shirai}, K.},
  \bibinfo{author}{{Sawada}, H.}, \bibinfo{author}{{Ishibashi}, K.},
  \bibinfo{author}{{Honda}, R.}, \bibinfo{author}{{Sakatani}, N.},
  \bibinfo{author}{{Iijima}, Y.}, \bibinfo{author}{{Okamoto}, C.},
  \bibinfo{author}{{Yano}, H.}, \bibinfo{author}{{Takagi}, Y.},
  \bibinfo{author}{{Hayakawa}, M.}, \bibinfo{author}{{Michel}, P.},
  \bibinfo{author}{{Jutzi}, M.}, \bibinfo{author}{{Shimaki}, Y.},
  \bibinfo{author}{{Kimura}, S.}, \bibinfo{author}{{Mimasu}, Y.},
  \bibinfo{author}{{Toda}, T.}, \bibinfo{author}{{Imamura}, H.},
  \bibinfo{author}{{Nakazawa}, S.}, \bibinfo{author}{{Hayakawa}, H.},
  \bibinfo{author}{{Sugita}, S.}, \bibinfo{author}{{Morota}, T.},
  \bibinfo{author}{{Kameda}, S.}, \bibinfo{author}{{Tatsumi}, E.},
  \bibinfo{author}{{Cho}, Y.}, \bibinfo{author}{{Yoshioka}, K.},
  \bibinfo{author}{{Yokota}, Y.}, \bibinfo{author}{{Matsuoka}, M.},
  \bibinfo{author}{{Yamada}, M.}, \bibinfo{author}{{Kouyama}, T.},
  \bibinfo{author}{{Honda}, C.}, \bibinfo{author}{{Tsuda}, Y.},
  \bibinfo{author}{{Watanabe}, S.}, \bibinfo{author}{{Yoshikawa}, M.},
  \bibinfo{author}{{Tanaka}, S.}, \bibinfo{author}{{Terui}, F.},
  \bibinfo{author}{{Kikuchi}, S.}, \bibinfo{author}{{Yamaguchi}, T.},
  \bibinfo{author}{{Ogawa}, N.}, \bibinfo{author}{{Ono}, G.},
  \bibinfo{author}{{Yoshikawa}, K.}, \bibinfo{author}{{Takahashi}, T.},
  \bibinfo{author}{{Takei}, Y.}, \bibinfo{author}{{Fujii}, A.},
  \bibinfo{author}{{Takeuchi}, H.}, \bibinfo{author}{{Yamamoto}, Y.},
  \bibinfo{author}{{Okada}, T.}, \bibinfo{author}{{Hirose}, C.},
  \bibinfo{author}{{Hosoda}, S.}, \bibinfo{author}{{Mori}, O.},
  \bibinfo{author}{{Shimada}, T.}, \bibinfo{author}{{Soldini}, S.},
  \bibinfo{author}{{Tsukizaki}, R.}, \bibinfo{author}{{Iwata}, T.},
  \bibinfo{author}{{Ozaki}, M.}, \bibinfo{author}{{Abe}, M.},
  \bibinfo{author}{{Namiki}, N.}, \bibinfo{author}{{Kitazato}, K.},
  \bibinfo{author}{{Tachibana}, S.}, \bibinfo{author}{{Ikeda}, H.},
  \bibinfo{author}{{Hirata}, N.}, \bibinfo{author}{{Hirata}, N.},
  \bibinfo{author}{{Noguchi}, R.}, \& \bibinfo{author}{{Miura}, A.}
  (\bibinfo{year}{2020}).
\newblock \bibinfo{title}{{An artificial impact on the asteroid (162173) Ryugu
  formed a crater in the gravity-dominated regime}}.
\newblock {\it \bibinfo{journal}{Science}\/},  {\it \bibinfo{volume}{368}\/},
  \bibinfo{pages}{67--71}. \DOIprefix\doi{10.1126/science.aaz1701}.
\bibitem[{{Avdellidou} et~al.(2020){Avdellidou}, {Di Donna}, {Schultz},
  {Harthong}, {Price}, {Peyroux}, {Britt}, {Cole} \& {Delbo'}}]{avdellidou2020}
\bibinfo{author}{{Avdellidou}, C.}, \bibinfo{author}{{Di Donna}, A.},
  \bibinfo{author}{{Schultz}, C.}, \bibinfo{author}{{Harthong}, B.},
  \bibinfo{author}{{Price}, M.~C.}, \bibinfo{author}{{Peyroux}, R.},
  \bibinfo{author}{{Britt}, D.}, \bibinfo{author}{{Cole}, M.}, \&
  \bibinfo{author}{{Delbo'}, M.} (\bibinfo{year}{2020}).
\newblock \bibinfo{title}{{Very weak carbonaceous asteroid simulants I:
  Mechanical properties and response to hypervelocity impacts}}.
\newblock {\it \bibinfo{journal}{\icarus}\/},  {\it \bibinfo{volume}{341}\/},
  \bibinfo{pages}{113648}. \DOIprefix\doi{10.1016/j.icarus.2020.113648}.
\bibitem[{{Avdellidou} et~al.(2017){Avdellidou}, {Price}, {Delbo} \&
  {Cole}}]{avdellidou2017}
\bibinfo{author}{{Avdellidou}, C.}, \bibinfo{author}{{Price}, M.~C.},
  \bibinfo{author}{{Delbo}, M.}, \& \bibinfo{author}{{Cole}, M.~J.}
  (\bibinfo{year}{2017}).
\newblock \bibinfo{title}{{Survival of the impactor during hypervelocity
  collisions - II. An analogue for high-porosity targets}}.
\newblock {\it \bibinfo{journal}{MNRAS}\/},  {\it \bibinfo{volume}{464}\/},
  \bibinfo{pages}{734--738}. \DOIprefix\doi{10.1093/mnras/stw2381}.
  \href{http://arxiv.org/abs/1612.05060}{\tt arXiv:1612.05060}.
\bibitem[{{Avdellidou} et~al.(2016){Avdellidou}, {Price}, {Delbo}, {Ioannidis}
  \& {Cole}}]{avdellidou2016}
\bibinfo{author}{{Avdellidou}, C.}, \bibinfo{author}{{Price}, M.~C.},
  \bibinfo{author}{{Delbo}, M.}, \bibinfo{author}{{Ioannidis}, P.}, \&
  \bibinfo{author}{{Cole}, M.~J.} (\bibinfo{year}{2016}).
\newblock \bibinfo{title}{{Survival of the impactor during hypervelocity
  collisions - I. An analogue for low porosity targets}}.
\newblock {\it \bibinfo{journal}{MNRAS}\/},  {\it \bibinfo{volume}{456}\/},
  \bibinfo{pages}{2957}. \DOIprefix\doi{10.1093/mnras/stv2844}.
  \href{http://arxiv.org/abs/1512.03262}{\tt arXiv:1512.03262}.
\bibitem[{{Avdellidou} \& {Vaubaillon}(2019)}]{avdellidou2019}
\bibinfo{author}{{Avdellidou}, C.}, \& \bibinfo{author}{{Vaubaillon}, J.}
  (\bibinfo{year}{2019}).
\newblock \bibinfo{title}{{Temperatures of lunar impact flashes: mass and size
  distribution of small impactors hitting the Moon}}.
\newblock {\it \bibinfo{journal}{\mnras}\/},  {\it \bibinfo{volume}{484}\/},
  \bibinfo{pages}{5212--5222}. \DOIprefix\doi{10.1093/mnras/stz355}.
  \href{http://arxiv.org/abs/1902.00987}{\tt arXiv:1902.00987}.
\bibitem[{{Babadzhanov} \& {Kokhirova}(2009)}]{babadzhanov2009}
\bibinfo{author}{{Babadzhanov}, P.~B.}, \& \bibinfo{author}{{Kokhirova}, G.~I.}
  (\bibinfo{year}{2009}).
\newblock \bibinfo{title}{{Densities and porosities of meteoroids}}.
\newblock {\it \bibinfo{journal}{\aap}\/},  {\it \bibinfo{volume}{495}\/},
  \bibinfo{pages}{353--358}. \DOIprefix\doi{10.1051/0004-6361:200810460}.
\bibitem[{{Baldwin}(1971)}]{baldwin1971}
\bibinfo{author}{{Baldwin}, R.~B.} (\bibinfo{year}{1971}).
\newblock \bibinfo{title}{{On the History of Lunar Impact Cratering: The
  Absolute Time Scale and the Origin of Planetesimals}}.
\newblock {\it \bibinfo{journal}{\icarus}\/},  {\it \bibinfo{volume}{14}\/},
  \bibinfo{pages}{36--52}. \DOIprefix\doi{10.1016/0019-1035(71)90100-X}.
\bibitem[{{Ballouz} et~al.(2020){Ballouz}, {Walsh}, {Barnouin},
  {DellaGiustina}, {Al Asad}, {Jawin}, {Daly}, {Bottke}, {Michel},
  {Avdellidou}, {Delbo}, {Daly}, {Asphaug}, {Bennett}, {Bierhaus}, {Connolly
  Jr}, {Golish}, {Molaro}, Nolan, {Pajola}, {Rizk}, {Schwartz}, {Trang} \&
  {Wolner}}]{ballouz2020}
\bibinfo{author}{{Ballouz}, R.~L.}, \bibinfo{author}{{Walsh}, K.~J.},
  \bibinfo{author}{{Barnouin}, O.~S.}, \bibinfo{author}{{DellaGiustina},
  D.~N.}, \bibinfo{author}{{Al Asad}, M.}, \bibinfo{author}{{Jawin}, E.~R.},
  \bibinfo{author}{{Daly}, M.~G.}, \bibinfo{author}{{Bottke}, B.~F.},
  \bibinfo{author}{{Michel}, P.}, \bibinfo{author}{{Avdellidou}, C.},
  \bibinfo{author}{{Delbo}, M.}, \bibinfo{author}{{Daly}, R.~T.},
  \bibinfo{author}{{Asphaug}, E.}, \bibinfo{author}{{Bennett}, C.~A.},
  \bibinfo{author}{{Bierhaus}, E.~B.}, \bibinfo{author}{{Connolly Jr}, H.~C.},
  \bibinfo{author}{{Golish}, D.~R.}, \bibinfo{author}{{Molaro}, J.~L.},
  \bibinfo{author}{Nolan, M.~C.}, \bibinfo{author}{{Pajola}, M.},
  \bibinfo{author}{{Rizk}, B.}, \bibinfo{author}{{Schwartz}, S.~R.},
  \bibinfo{author}{{Trang}, D.}, \& \bibinfo{author}{{Wolner}, D.~S., C.~W.~V.
  \&~{Lauretta}} (\bibinfo{year}{2020}).
\newblock \bibinfo{title}{{Bennu’s near-Earth lifetime of 1.75 million years
  inferred from craters on its boulders}}.
\newblock {\it \bibinfo{journal}{Science}\/},  {\it \bibinfo{volume}{587}\/},
  \bibinfo{pages}{205--209}.
  \DOIprefix\doi{https://doi.org/10.1038/s41586-020-2846-z}.
\bibitem[{{Bellot Rubio} et~al.(2000){Bellot Rubio}, {Ortiz} \&
  {Sada}}]{bellotrubio2000}
\bibinfo{author}{{Bellot Rubio}, L.~R.}, \bibinfo{author}{{Ortiz}, J.~L.}, \&
  \bibinfo{author}{{Sada}, P.~V.} (\bibinfo{year}{2000}).
\newblock \bibinfo{title}{{Observation and Interpretation of Meteoroid Impact
  Flashes on the Moon}}.
\newblock {\it \bibinfo{journal}{Earth Moon and Planets}\/},  {\it
  \bibinfo{volume}{82}\/}, \bibinfo{pages}{575--598}.
\bibitem[{{Bertin} \& {Arnouts}(1996)}]{1996A&AS..117..393B}
\bibinfo{author}{{Bertin}, E.}, \& \bibinfo{author}{{Arnouts}, S.}
  (\bibinfo{year}{1996}).
\newblock \bibinfo{title}{{SExtractor: Software for source extraction.}}
\newblock {\it \bibinfo{journal}{\aap}\/},  {\it \bibinfo{volume}{117}\/},
  \bibinfo{pages}{393}. \DOIprefix\doi{10.1051/aas:1996164}.
\bibitem[{{Bessell} et~al.(1998){Bessell}, {Castelli} \& {Plez}}]{bessell1998}
\bibinfo{author}{{Bessell}, M.~S.}, \bibinfo{author}{{Castelli}, F.}, \&
  \bibinfo{author}{{Plez}, B.} (\bibinfo{year}{1998}).
\newblock \bibinfo{title}{{Model atmospheres broad-band colors, bolometric
  corrections and temperature calibrations for O - M stars}}.
\newblock {\it \bibinfo{journal}{\aap}\/},  {\it \bibinfo{volume}{333}\/},
  \bibinfo{pages}{231--250}.
\bibitem[{{Bonanos} et~al.(2018){Bonanos}, {Avdellidou}, {Liakos}, {Xilouris},
  {Dapergolas}, {Koschny}, {Bellas- Velidis}, {Boumis}, {Charmandaris},
  {Fytsilis} \& {Maroussis}}]{bonanos2018}
\bibinfo{author}{{Bonanos}, A.~Z.}, \bibinfo{author}{{Avdellidou}, C.},
  \bibinfo{author}{{Liakos}, A.}, \bibinfo{author}{{Xilouris}, E.~M.},
  \bibinfo{author}{{Dapergolas}, A.}, \bibinfo{author}{{Koschny}, D.},
  \bibinfo{author}{{Bellas- Velidis}, I.}, \bibinfo{author}{{Boumis}, P.},
  \bibinfo{author}{{Charmandaris}, V.}, \bibinfo{author}{{Fytsilis}, A.}, \&
  \bibinfo{author}{{Maroussis}, A.} (\bibinfo{year}{2018}).
\newblock \bibinfo{title}{{NELIOTA: First temperature measurement of lunar
  impact flashes}}.
\newblock {\it \bibinfo{journal}{\aap}\/},  {\it \bibinfo{volume}{612}\/}.
  \DOIprefix\doi{10.1051/0004-6361/201732109}.
\bibitem[{{Bottke} et~al.(2002){Bottke}, {Morbidelli}, {Jedicke}, {Petit},
  {Levison}, {Michel} \& {Metcalfe}}]{bottke2002}
\bibinfo{author}{{Bottke}, W.~F.}, \bibinfo{author}{{Morbidelli}, A.},
  \bibinfo{author}{{Jedicke}, R.}, \bibinfo{author}{{Petit}, J.-M.},
  \bibinfo{author}{{Levison}, H.~F.}, \bibinfo{author}{{Michel}, P.}, \&
  \bibinfo{author}{{Metcalfe}, T.~S.} (\bibinfo{year}{2002}).
\newblock \bibinfo{title}{{Debiased Orbital and Absolute Magnitude Distribution
  of the Near-Earth Objects}}.
\newblock {\it \bibinfo{journal}{Icarus}\/},  {\it \bibinfo{volume}{156}\/},
  \bibinfo{pages}{399--433}. \DOIprefix\doi{10.1006/icar.2001.6788}.
\bibitem[{{Bouley} et~al.(2012){Bouley}, {Baratoux}, {Vaubaillon}, {Mocquet},
  {Le Feuvre}, {Colas}, {Benkhaldoun}, {Daassou}, {Sabil} \&
  {Lognonn{\'e}}}]{bouley2012}
\bibinfo{author}{{Bouley}, S.}, \bibinfo{author}{{Baratoux}, D.},
  \bibinfo{author}{{Vaubaillon}, J.}, \bibinfo{author}{{Mocquet}, A.},
  \bibinfo{author}{{Le Feuvre}, M.}, \bibinfo{author}{{Colas}, F.},
  \bibinfo{author}{{Benkhaldoun}, Z.}, \bibinfo{author}{{Daassou}, A.},
  \bibinfo{author}{{Sabil}, M.}, \& \bibinfo{author}{{Lognonn{\'e}}, P.}
  (\bibinfo{year}{2012}).
\newblock \bibinfo{title}{{Power and duration of impact flashes on the Moon:
  Implication for the cause of radiation}}.
\newblock {\it \bibinfo{journal}{Icarus}\/},  {\it \bibinfo{volume}{218}\/},
  \bibinfo{pages}{115--124}. \DOIprefix\doi{10.1016/j.icarus.2011.11.028}.
\bibitem[{{Burchell} et~al.(2015){Burchell}, {Cole}, {Ramkissoon},
  {Wozniakiewicz}, {Price} \& {Foing}}]{burchell2015}
\bibinfo{author}{{Burchell}, M.~J.}, \bibinfo{author}{{Cole}, M.~J.},
  \bibinfo{author}{{Ramkissoon}, N.~K.}, \bibinfo{author}{{Wozniakiewicz},
  P.~J.}, \bibinfo{author}{{Price}, M.~C.}, \& \bibinfo{author}{{Foing}, B.}
  (\bibinfo{year}{2015}).
\newblock \bibinfo{title}{{SMART-1 end of life shallow regolith impact
  simulations}}.
\newblock {\it \bibinfo{journal}{Meteoritics and Planetary Science}\/},  {\it
  \bibinfo{volume}{50}\/}, \bibinfo{pages}{1436--1448}.
  \DOIprefix\doi{10.1111/maps.12479}.
\bibitem[{{Cheng} et~al.(2016){Cheng}, {Michel}, {Jutzi}, {Rivkin}, {Stickle},
  {Barnouin}, {Ernst}, {Atchison}, {Pravec}, {Richardson} \& {AIDA
  Team}}]{cheng2016}
\bibinfo{author}{{Cheng}, A.~F.}, \bibinfo{author}{{Michel}, P.},
  \bibinfo{author}{{Jutzi}, M.}, \bibinfo{author}{{Rivkin}, A.~S.},
  \bibinfo{author}{{Stickle}, A.}, \bibinfo{author}{{Barnouin}, O.},
  \bibinfo{author}{{Ernst}, C.}, \bibinfo{author}{{Atchison}, J.},
  \bibinfo{author}{{Pravec}, P.}, \bibinfo{author}{{Richardson}, D.~C.}, \&
  \bibinfo{author}{{AIDA Team}} (\bibinfo{year}{2016}).
\newblock \bibinfo{title}{{Asteroid Impact \& Deflection Assessment mission:
  Kinetic impactor}}.
\newblock {\it \bibinfo{journal}{\planss}\/},  {\it \bibinfo{volume}{121}\/},
  \bibinfo{pages}{27--35}. \DOIprefix\doi{10.1016/j.pss.2015.12.004}.
\bibitem[{{Cintala}(1992)}]{cintala1992}
\bibinfo{author}{{Cintala}, M.~J.} (\bibinfo{year}{1992}).
\newblock \bibinfo{title}{{Impact-induced thermal effects in the lunar and
  Mercurian regoliths}}.
\newblock {\it \bibinfo{journal}{\jgr}\/},  {\it \bibinfo{volume}{97}\/},
  \bibinfo{pages}{947--973}. \DOIprefix\doi{10.1029/91JE02207}.
\bibitem[{{Colaprete} et~al.(2010){Colaprete}, {Schultz}, {Heldmann}, {Wooden},
  {Shirley}, {Ennico}, {Hermalyn}, {Marshall}, {Ricco}, {Elphic}, {Goldstein},
  {Summy}, {Bart}, {Asphaug}, {Korycansky}, {Landis} \&
  {Sollitt}}]{colaprete2010}
\bibinfo{author}{{Colaprete}, A.}, \bibinfo{author}{{Schultz}, P.},
  \bibinfo{author}{{Heldmann}, J.}, \bibinfo{author}{{Wooden}, D.},
  \bibinfo{author}{{Shirley}, M.}, \bibinfo{author}{{Ennico}, K.},
  \bibinfo{author}{{Hermalyn}, B.}, \bibinfo{author}{{Marshall}, W.},
  \bibinfo{author}{{Ricco}, A.}, \bibinfo{author}{{Elphic}, R.~C.},
  \bibinfo{author}{{Goldstein}, D.}, \bibinfo{author}{{Summy}, D.},
  \bibinfo{author}{{Bart}, G.~D.}, \bibinfo{author}{{Asphaug}, E.},
  \bibinfo{author}{{Korycansky}, D.}, \bibinfo{author}{{Landis}, D.}, \&
  \bibinfo{author}{{Sollitt}, L.} (\bibinfo{year}{2010}).
\newblock \bibinfo{title}{{Detection of Water in the LCROSS Ejecta Plume}}.
\newblock {\it \bibinfo{journal}{Science}\/},  {\it \bibinfo{volume}{330}\/},
  \bibinfo{pages}{463}. \DOIprefix\doi{10.1126/science.1186986}.
\bibitem[{{Cudnik}(2009)}]{cudnik2009}
\bibinfo{author}{{Cudnik}, B.} (\bibinfo{year}{2009}).
\newblock {\it \bibinfo{title}{{Lunar Meteoroid Impacts and How to Observe
  Them}}\/}.
\newblock \DOIprefix\doi{10.1007/978-1-4419-0324-2}.
\bibitem[{{Cudnik} et~al.(2003){Cudnik}, {Dunham}, {Palmer}, {Cook}, {Venable}
  \& {Gural}}]{cudnik2003}
\bibinfo{author}{{Cudnik}, B.~M.}, \bibinfo{author}{{Dunham}, D.~W.},
  \bibinfo{author}{{Palmer}, D.~M.}, \bibinfo{author}{{Cook}, A.},
  \bibinfo{author}{{Venable}, R.}, \& \bibinfo{author}{{Gural}, P.~S.}
  (\bibinfo{year}{2003}).
\newblock \bibinfo{title}{{Ground-Based Observations Of Lunar Meteoritic
  Phenomena}}.
\newblock {\it \bibinfo{journal}{Earth Moon and Planets}\/},  {\it
  \bibinfo{volume}{93}\/}, \bibinfo{pages}{145--161}.
  \DOIprefix\doi{10.1023/B:MOON.0000047475.61749.c1}.
\bibitem[{{Daly} \& {Schultz}(2015)}]{daly2015}
\bibinfo{author}{{Daly}, R.~T.}, \& \bibinfo{author}{{Schultz}, P.~H.}
  (\bibinfo{year}{2015}).
\newblock \bibinfo{title}{{New Constraints on the Delivery of Impactors to Icy
  Bodies: Implications for Ceres}}.
\newblock In {\it \bibinfo{booktitle}{Lunar and Planetary Science
  Conference}\/} (p. \bibinfo{pages}{1972}).
\newblock volume~\bibinfo{volume}{46}.
\bibitem[{{Daly} \& {Schultz}(2016)}]{daly2016}
\bibinfo{author}{{Daly}, R.~T.}, \& \bibinfo{author}{{Schultz}, P.~H.}
  (\bibinfo{year}{2016}).
\newblock \bibinfo{title}{{Delivering a projectile component to the vestan
  regolith}}.
\newblock {\it \bibinfo{journal}{Icarus}\/},  {\it \bibinfo{volume}{264}\/},
  \bibinfo{pages}{9}. \DOIprefix\doi{10.1016/j.icarus.2015.08.034}.
\bibitem[{{Dunham} et~al.(2000){Dunham}, {Sterner}, {Gotwols}, {Cudnik},
  {Palmer}, {Sada} \& {Frankenberger}}]{dunham2000}
\bibinfo{author}{{Dunham}, D.~W.}, \bibinfo{author}{{Sterner}, I., Ray},
  \bibinfo{author}{{Gotwols}, B.}, \bibinfo{author}{{Cudnik}, B.~M.},
  \bibinfo{author}{{Palmer}, D.~M.}, \bibinfo{author}{{Sada}, P.~V.}, \&
  \bibinfo{author}{{Frankenberger}, R.} (\bibinfo{year}{2000}).
\newblock \bibinfo{title}{{Confirmed lunar meteor impacts from the November
  1999 Leonids}}.
\newblock {\it \bibinfo{journal}{Occultation Newsletter, International
  Occultation Timing Association (IOTA)}\/},  {\it \bibinfo{volume}{8}\/},
  \bibinfo{pages}{9--11}.
\bibitem[{{Eichhorn}(1975)}]{eichhorn1975}
\bibinfo{author}{{Eichhorn}, G.} (\bibinfo{year}{1975}).
\newblock \bibinfo{title}{{Measurements of the light flash produced by high
  velocity particle impact}}.
\newblock {\it \bibinfo{journal}{\planss}\/},  {\it \bibinfo{volume}{23}\/},
  \bibinfo{pages}{1519--1525}. \DOIprefix\doi{10.1016/0032-0633(75)90005-7}.
\bibitem[{{Eichhorn}(1976)}]{eichhorn1976}
\bibinfo{author}{{Eichhorn}, G.} (\bibinfo{year}{1976}).
\newblock \bibinfo{title}{{Analysis of the hypervelocity impact process from
  impact flash measurements}}.
\newblock {\it \bibinfo{journal}{\planss}\/},  {\it \bibinfo{volume}{24}\/},
  \bibinfo{pages}{771--781}. \DOIprefix\doi{10.1016/0032-0633(76)90114-8}.
\bibitem[{{Flynn} et~al.(2020){Flynn}, {Durda}, {Molesky}, {May}, {Congram},
  {Loftus}, {Reagan}, {Strait} \& {Macke}}]{flynn2020}
\bibinfo{author}{{Flynn}, G.~J.}, \bibinfo{author}{{Durda}, D.~D.},
  \bibinfo{author}{{Molesky}, M.~J.}, \bibinfo{author}{{May}, B.~A.},
  \bibinfo{author}{{Congram}, S.~N.}, \bibinfo{author}{{Loftus}, C.~L.},
  \bibinfo{author}{{Reagan}, J.~R.}, \bibinfo{author}{{Strait}, M.~M.}, \&
  \bibinfo{author}{{Macke}, R.~J.} (\bibinfo{year}{2020}).
\newblock \bibinfo{title}{{Hypervelocity cratering and disruption of the
  Northwest Africa 4502 carbonaceous chondrite meteorite: Implications for
  crater production, catastrophic disruption, momentum transfer and dust
  production on asteroids}}.
\newblock {\it \bibinfo{journal}{\planss}\/},  {\it \bibinfo{volume}{187}\/},
  \bibinfo{pages}{104916}. \DOIprefix\doi{10.1016/j.pss.2020.104916}.
\bibitem[{{Flynn} et~al.(2018){Flynn}, {Durda}, {Patmore}, {Jack}, {Molesky},
  {May}, {Congram}, {Strait} \& {Macke}}]{flynn2018}
\bibinfo{author}{{Flynn}, G.~J.}, \bibinfo{author}{{Durda}, D.~D.},
  \bibinfo{author}{{Patmore}, E.~B.}, \bibinfo{author}{{Jack}, S.~J.},
  \bibinfo{author}{{Molesky}, M.~J.}, \bibinfo{author}{{May}, B.~A.},
  \bibinfo{author}{{Congram}, S.~N.}, \bibinfo{author}{{Strait}, M.~M.}, \&
  \bibinfo{author}{{Macke}, R.~J.} (\bibinfo{year}{2018}).
\newblock \bibinfo{title}{{Hypervelocity cratering and disruption of the
  Northwest Africa 869 ordinary chondrite meteorite: Implications for crater
  production, catastrophic disruption, momentum transfer and dust production on
  asteroids}}.
\newblock {\it \bibinfo{journal}{\planss}\/},  {\it \bibinfo{volume}{164}\/},
  \bibinfo{pages}{91--105}. \DOIprefix\doi{10.1016/j.pss.2018.06.019}.
\bibitem[{{Foing} et~al.(2018){Foing}, {Racca}, {Marini}, {Koschny}, {Frew},
  {Grieger}, {Camino-Ramos}, {Josset}, {Grande}, {Smart-1 Science} \&
  {Technology Working Team}}]{foing2018}
\bibinfo{author}{{Foing}, B.~H.}, \bibinfo{author}{{Racca}, G.},
  \bibinfo{author}{{Marini}, A.}, \bibinfo{author}{{Koschny}, D.},
  \bibinfo{author}{{Frew}, D.}, \bibinfo{author}{{Grieger}, B.},
  \bibinfo{author}{{Camino-Ramos}, O.}, \bibinfo{author}{{Josset}, J.~L.},
  \bibinfo{author}{{Grande}, M.}, \bibinfo{author}{{Smart-1 Science}}, \&
  \bibinfo{author}{{Technology Working Team}} (\bibinfo{year}{2018}).
\newblock \bibinfo{title}{{SMART-1 technology, scientific results and heritage
  for future space missions}}.
\newblock {\it \bibinfo{journal}{\planss}\/},  {\it \bibinfo{volume}{151}\/},
  \bibinfo{pages}{141--148}. \DOIprefix\doi{10.1016/j.pss.2017.09.002}.
\bibitem[{{Halliday} et~al.(1996){Halliday}, {Griffin} \&
  {Blackwell}}]{halliday1996}
\bibinfo{author}{{Halliday}, I.}, \bibinfo{author}{{Griffin}, A.~A.}, \&
  \bibinfo{author}{{Blackwell}, A.~T.} (\bibinfo{year}{1996}).
\newblock \bibinfo{title}{{Detailed data for 259 fireballs from the Canadian
  camera network and inferences concerning the influx of large meteoroids}}.
\newblock {\it \bibinfo{journal}{Meteoritics and Planetary Science}\/},  {\it
  \bibinfo{volume}{31}\/}, \bibinfo{pages}{185--217}.
  \DOIprefix\doi{10.1111/j.1945-5100.1996.tb02014.x}.
\bibitem[{{Hartmann} \& {Neukum}(2001)}]{hartmann2001}
\bibinfo{author}{{Hartmann}, W.~K.}, \& \bibinfo{author}{{Neukum}, G.}
  (\bibinfo{year}{2001}).
\newblock \bibinfo{title}{{Cratering Chronology and the Evolution of Mars}}.
\newblock {\it \bibinfo{journal}{\ssr}\/},  {\it \bibinfo{volume}{96}\/},
  \bibinfo{pages}{165--194}. \DOIprefix\doi{10.1023/A:1011945222010}.
\bibitem[{{Hayne} et~al.(2010){Hayne}, {Greenhagen}, {Foote}, {Siegler},
  {Vasavada} \& {Paige}}]{hayne2010}
\bibinfo{author}{{Hayne}, P.~O.}, \bibinfo{author}{{Greenhagen}, B.~T.},
  \bibinfo{author}{{Foote}, M.~C.}, \bibinfo{author}{{Siegler}, M.~A.},
  \bibinfo{author}{{Vasavada}, A.~R.}, \& \bibinfo{author}{{Paige}, D.~A.}
  (\bibinfo{year}{2010}).
\newblock \bibinfo{title}{{Diviner Lunar Radiometer Observations of the LCROSS
  Impact}}.
\newblock {\it \bibinfo{journal}{Science}\/},  {\it \bibinfo{volume}{330}\/},
  \bibinfo{pages}{477}. \DOIprefix\doi{10.1126/science.1197135}.
\bibitem[{{Holsapple}(1994)}]{holsapple1994}
\bibinfo{author}{{Holsapple}, K.~A.} (\bibinfo{year}{1994}).
\newblock \bibinfo{title}{{Catastrophic disruptions and cratering of solar
  system bodies: a review and new results}}.
\newblock {\it \bibinfo{journal}{\planss}\/},  {\it \bibinfo{volume}{42}\/},
  \bibinfo{pages}{1067--1078}. \DOIprefix\doi{10.1016/0032-0633(94)90007-8}.
\bibitem[{{Hughes}(1987)}]{hughes1987}
\bibinfo{author}{{Hughes}, D.~W.} (\bibinfo{year}{1987}).
\newblock \bibinfo{title}{{P/Halley dust characteristics - A comparison between
  Orionid and Eta Aquarid meteor observations and those from the flyby
  spacecraft}}.
\newblock {\it \bibinfo{journal}{\aap}\/},  {\it \bibinfo{volume}{187}\/},
  \bibinfo{pages}{879--888}.
\bibitem[{{Ivanov}(2001)}]{ivanov2001}
\bibinfo{author}{{Ivanov}, B.~A.} (\bibinfo{year}{2001}).
\newblock \bibinfo{title}{{Mars/Moon Cratering Rate Ratio Estimates}}.
\newblock {\it \bibinfo{journal}{\ssr}\/},  {\it \bibinfo{volume}{96}\/},
  \bibinfo{pages}{87--104}.
\bibitem[{{Jenniskens}(1994)}]{jenniskens1994}
\bibinfo{author}{{Jenniskens}, P.} (\bibinfo{year}{1994}).
\newblock \bibinfo{title}{{Meteor stream activity I. The annual streams}}.
\newblock {\it \bibinfo{journal}{\aap}\/},  {\it \bibinfo{volume}{287}\/},
  \bibinfo{pages}{990--1013}.
\bibitem[{{Jenniskens}(2006)}]{jenniskens2006}
\bibinfo{author}{{Jenniskens}, P.} (\bibinfo{year}{2006}).
\newblock {\it \bibinfo{title}{{Meteor Showers and their Parent Comets}}\/}.
\bibitem[{{Kopal}(1965)}]{Kopal1965}
\bibinfo{author}{{Kopal}, Z.} (\bibinfo{year}{1965}).
\newblock \bibinfo{title}{{Topography of the Moon}}.
\newblock {\it \bibinfo{journal}{\ssr}\/},  {\it \bibinfo{volume}{4}\/},
  \bibinfo{pages}{737--855}. \DOIprefix\doi{10.1007/BF00216275}.
\bibitem[{{Korycansky} \& {Zahnle}(2005)}]{korycansky2005}
\bibinfo{author}{{Korycansky}, D.~G.}, \& \bibinfo{author}{{Zahnle}, K.~J.}
  (\bibinfo{year}{2005}).
\newblock \bibinfo{title}{{Modeling crater populations on Venus and Titan}}.
\newblock {\it \bibinfo{journal}{\planss}\/},  {\it \bibinfo{volume}{53}\/},
  \bibinfo{pages}{695--710}. \DOIprefix\doi{10.1016/j.pss.2005.03.002}.
\bibitem[{{Larson} et~al.(2019){Larson}, {Hayne} \& {Avdellidou}}]{larson2019}
\bibinfo{author}{{Larson}, R.}, \bibinfo{author}{{Hayne}, P.}, \&
  \bibinfo{author}{{Avdellidou}, C.} (\bibinfo{year}{2019}).
\newblock \bibinfo{title}{{Automating the detection and coordinate
  identification of impact flashes on the Lunar surface}}.
\newblock In {\it \bibinfo{booktitle}{EPSC-DPS Joint Meeting 2019}\/} (pp.
  \bibinfo{pages}{EPSC--DPS2019--1193}).
\newblock volume \bibinfo{volume}{2019}.
\bibitem[{{Le Feuvre} \& {Wieczorek}(2011)}]{lefeuvre2011}
\bibinfo{author}{{Le Feuvre}, M.}, \& \bibinfo{author}{{Wieczorek}, M.~A.}
  (\bibinfo{year}{2011}).
\newblock \bibinfo{title}{{Nonuniform cratering of the Moon and a revised
  crater chronology of the inner Solar System}}.
\newblock {\it \bibinfo{journal}{\icarus}\/},  {\it \bibinfo{volume}{214}\/},
  \bibinfo{pages}{1--20}. \DOIprefix\doi{10.1016/j.icarus.2011.03.010}.
\bibitem[{{Liakos} et~al.(2020){Liakos}, {Bonanos}, {Xilouris}, {Koschny},
  {Bellas-Velidis}, {Boumis}, {Charmand aris}, {Dapergolas}, {Fytsilis},
  {Maroussis} \& {Moissl}}]{liakos2020}
\bibinfo{author}{{Liakos}, A.}, \bibinfo{author}{{Bonanos}, A.~Z.},
  \bibinfo{author}{{Xilouris}, E.~M.}, \bibinfo{author}{{Koschny}, D.},
  \bibinfo{author}{{Bellas-Velidis}, I.}, \bibinfo{author}{{Boumis}, P.},
  \bibinfo{author}{{Charmand aris}, V.}, \bibinfo{author}{{Dapergolas}, A.},
  \bibinfo{author}{{Fytsilis}, A.}, \bibinfo{author}{{Maroussis}, A.}, \&
  \bibinfo{author}{{Moissl}, R.} (\bibinfo{year}{2020}).
\newblock \bibinfo{title}{{NELIOTA: Methods, statistics, and results for
  meteoroids impacting the Moon}}.
\newblock {\it \bibinfo{journal}{\aap}\/},  {\it \bibinfo{volume}{633}\/},
  \bibinfo{pages}{A112}. \DOIprefix\doi{10.1051/0004-6361/201936709}.
  \href{http://arxiv.org/abs/1911.06101}{\tt arXiv:1911.06101}.
\bibitem[{{Madiedo} et~al.(2018){Madiedo}, {Ortiz} \& {Morales}}]{madiedo2018}
\bibinfo{author}{{Madiedo}, J.~M.}, \bibinfo{author}{{Ortiz}, J.~L.}, \&
  \bibinfo{author}{{Morales}, N.} (\bibinfo{year}{2018}).
\newblock \bibinfo{title}{{The first observations to determine the temperature
  of a lunar impact flash and its evolution}}.
\newblock {\it \bibinfo{journal}{\mnras}\/},  {\it \bibinfo{volume}{480}\/},
  \bibinfo{pages}{5010--5016}. \DOIprefix\doi{10.1093/mnras/sty1862}.
  \href{http://arxiv.org/abs/1807.03193}{\tt arXiv:1807.03193}.
\bibitem[{{Madiedo} et~al.(2014){Madiedo}, {Ortiz}, {Morales} \&
  {Cabrera-Ca{\~n}o}}]{madiedo2014}
\bibinfo{author}{{Madiedo}, J.~M.}, \bibinfo{author}{{Ortiz}, J.~L.},
  \bibinfo{author}{{Morales}, N.}, \& \bibinfo{author}{{Cabrera-Ca{\~n}o}, J.}
  (\bibinfo{year}{2014}).
\newblock \bibinfo{title}{{A large lunar impact blast on 2013 September 11}}.
\newblock {\it \bibinfo{journal}{MNRAS}\/},  {\it \bibinfo{volume}{439}\/},
  \bibinfo{pages}{2364--2369}. \DOIprefix\doi{10.1093/mnras/stu083}.
  \href{http://arxiv.org/abs/1402.5490}{\tt arXiv:1402.5490}.
\bibitem[{{Madiedo} et~al.(2015{\natexlab{a}}){Madiedo}, {Ortiz}, {Morales} \&
  {Cabrera-Ca{\~n}o}}]{madiedo2015b}
\bibinfo{author}{{Madiedo}, J.~M.}, \bibinfo{author}{{Ortiz}, J.~L.},
  \bibinfo{author}{{Morales}, N.}, \& \bibinfo{author}{{Cabrera-Ca{\~n}o}, J.}
  (\bibinfo{year}{2015}{\natexlab{a}}).
\newblock \bibinfo{title}{{MIDAS: Software for the detection and analysis of
  lunar impact flashes}}.
\newblock {\it \bibinfo{journal}{\planss}\/},  {\it \bibinfo{volume}{111}\/},
  \bibinfo{pages}{105--115}. \DOIprefix\doi{10.1016/j.pss.2015.03.018}.
  \href{http://arxiv.org/abs/1503.07018}{\tt arXiv:1503.07018}.
\bibitem[{{Madiedo} et~al.(2016){Madiedo}, {Ortiz}, {Morales} \&
  {Cabrera-Ca{\~n}o}}]{madiedo2016}
\bibinfo{author}{{Madiedo}, J.~M.}, \bibinfo{author}{{Ortiz}, J.~L.},
  \bibinfo{author}{{Morales}, N.}, \& \bibinfo{author}{{Cabrera-Ca{\~n}o}, J.}
  (\bibinfo{year}{2016}).
\newblock \bibinfo{title}{{Analysis of Lunar Impact Flashes Recorded During the
  Activity Period of the Lyrid Meteor Shower in 2013}}.
\newblock In {\it \bibinfo{booktitle}{Lunar and Planetary Science
  Conference}\/} Lunar and Planetary Science Conference (p.
  \bibinfo{pages}{1124}).
\bibitem[{{Madiedo} et~al.(2015{\natexlab{b}}){Madiedo}, {Ortiz}, {Organero},
  {Ana-Hern{\'a}ndez}, {Fonseca}, {Morales} \&
  {Cabrera-Ca{\~n}o}}]{madiedo2015}
\bibinfo{author}{{Madiedo}, J.~M.}, \bibinfo{author}{{Ortiz}, J.~L.},
  \bibinfo{author}{{Organero}, F.}, \bibinfo{author}{{Ana-Hern{\'a}ndez}, L.},
  \bibinfo{author}{{Fonseca}, F.}, \bibinfo{author}{{Morales}, N.}, \&
  \bibinfo{author}{{Cabrera-Ca{\~n}o}, J.}
  (\bibinfo{year}{2015}{\natexlab{b}}).
\newblock \bibinfo{title}{{Analysis of Moon impact flashes detected during the
  2012 and 2013 Perseids}}.
\newblock {\it \bibinfo{journal}{\aap}\/},  {\it \bibinfo{volume}{577}\/},
  \bibinfo{pages}{A118}. \DOIprefix\doi{10.1051/0004-6361/201525656}.
  \href{http://arxiv.org/abs/1503.05227}{\tt arXiv:1503.05227}.
\bibitem[{{Marchi} et~al.(2015){Marchi}, {Chapman}, {Barnouin}, {Richardson} \&
  {Vincent}}]{marchi2015}
\bibinfo{author}{{Marchi}, S.}, \bibinfo{author}{{Chapman}, C.~R.},
  \bibinfo{author}{{Barnouin}, O.~S.}, \bibinfo{author}{{Richardson}, J.~E.},
  \& \bibinfo{author}{{Vincent}, J.~B.} (\bibinfo{year}{2015}).
\newblock \bibinfo{title}{{Cratering on Asteroids}}.
\newblock In {\it \bibinfo{booktitle}{Asteroids IV}\/} (pp.
  \bibinfo{pages}{725--744}).
\newblock \DOIprefix\doi{10.2458/azu_uapress_9780816532131-ch037}.
\bibitem[{{Marchi} et~al.(2012){Marchi}, {Massironi}, {Vincent}, {Morbidelli},
  {Mottola}, {Marzari}, {K{\"u}ppers}, {Besse}, {Thomas}, {Barbieri}, {Naletto}
  \& {Sierks}}]{marchi2012}
\bibinfo{author}{{Marchi}, S.}, \bibinfo{author}{{Massironi}, M.},
  \bibinfo{author}{{Vincent}, J.~B.}, \bibinfo{author}{{Morbidelli}, A.},
  \bibinfo{author}{{Mottola}, S.}, \bibinfo{author}{{Marzari}, F.},
  \bibinfo{author}{{K{\"u}ppers}, M.}, \bibinfo{author}{{Besse}, S.},
  \bibinfo{author}{{Thomas}, N.}, \bibinfo{author}{{Barbieri}, C.},
  \bibinfo{author}{{Naletto}, G.}, \& \bibinfo{author}{{Sierks}, H.}
  (\bibinfo{year}{2012}).
\newblock \bibinfo{title}{{The cratering history of asteroid (21) Lutetia}}.
\newblock {\it \bibinfo{journal}{\planss}\/},  {\it \bibinfo{volume}{66}\/},
  \bibinfo{pages}{87--95}. \DOIprefix\doi{10.1016/j.pss.2011.10.010}.
  \href{http://arxiv.org/abs/1111.3628}{\tt arXiv:1111.3628}.
\bibitem[{{McNamara} et~al.(2004){McNamara}, {Jones}, {Kauffman}, {Suggs},
  {Cooke} \& {Smith}}]{mcnamara2004}
\bibinfo{author}{{McNamara}, H.}, \bibinfo{author}{{Jones}, J.},
  \bibinfo{author}{{Kauffman}, B.}, \bibinfo{author}{{Suggs}, R.},
  \bibinfo{author}{{Cooke}, W.}, \& \bibinfo{author}{{Smith}, S.}
  (\bibinfo{year}{2004}).
\newblock \bibinfo{title}{{Meteoroid Engineering Model (MEM): A Meteoroid Model
  For The Inner Solar System}}.
\newblock {\it \bibinfo{journal}{Earth Moon and Planets}\/},  {\it
  \bibinfo{volume}{95}\/}, \bibinfo{pages}{123--139}.
  \DOIprefix\doi{10.1007/s11038-005-9044-8}.
\bibitem[{{Moorhead} et~al.(2019){Moorhead}, {Egal}, {Brown}, {Moser} \&
  {Cooke}}]{moorhead2019}
\bibinfo{author}{{Moorhead}, A.~V.}, \bibinfo{author}{{Egal}, A.},
  \bibinfo{author}{{Brown}, P.~G.}, \bibinfo{author}{{Moser}, D.~E.}, \&
  \bibinfo{author}{{Cooke}, W.~J.} (\bibinfo{year}{2019}).
\newblock \bibinfo{title}{{Meteor shower forecasting in near-Earth space}}.
\newblock {\it \bibinfo{journal}{Journal of Spacecraft and Rockets}\/},  {\it
  \bibinfo{volume}{56}\/}, \bibinfo{pages}{1531--1545}.
  \DOIprefix\doi{10.2514/1.A34416}. \href{http://arxiv.org/abs/1904.06370}{\tt
  arXiv:1904.06370}.
\bibitem[{{Moser} et~al.(2011){Moser}, {Suggs}, {Swift}, {Suggs}, {Cooke},
  {Diekmann} \& {Koehler}}]{moser2011}
\bibinfo{author}{{Moser}, D.~E.}, \bibinfo{author}{{Suggs}, R.~M.},
  \bibinfo{author}{{Swift}, W.~R.}, \bibinfo{author}{{Suggs}, R.~J.},
  \bibinfo{author}{{Cooke}, W.~J.}, \bibinfo{author}{{Diekmann}, A.~M.}, \&
  \bibinfo{author}{{Koehler}, H.~M.} (\bibinfo{year}{2011}).
\newblock \bibinfo{title}{{Luminous Efficiency of Hypervelocity Meteoroid
  Impacts on the Moon Derived From the 2006 Geminids, 2007 Lyrids, and 2008
  Taurids}}.
\newblock In \bibinfo{editor}{W.~J. {Cooke}}, \bibinfo{editor}{D.~E. {Moser}},
  \bibinfo{editor}{B.~F. {Hardin}}, \& \bibinfo{editor}{D.~{Janches}} (Eds.),
  {\it \bibinfo{booktitle}{Meteoroids: The Smallest Solar System Bodies}\/}
  (pp. \bibinfo{pages}{142--154}).
\bibitem[{{Nemtchinov} et~al.(1998){Nemtchinov}, {Shuvalov}, {Artemieva},
  {Ivanov}, {Kosarev} \& {Trubetskaya}}]{nemtchinov1998}
\bibinfo{author}{{Nemtchinov}, I.~V.}, \bibinfo{author}{{Shuvalov}, V.~V.},
  \bibinfo{author}{{Artemieva}, N.~A.}, \bibinfo{author}{{Ivanov}, B.~A.},
  \bibinfo{author}{{Kosarev}, I.~B.}, \& \bibinfo{author}{{Trubetskaya}, I.~A.}
  (\bibinfo{year}{1998}).
\newblock \bibinfo{title}{{Light Impulse Created by Meteoroids Impacting the
  Moon}}.
\newblock In {\it \bibinfo{booktitle}{Lunar and Planetary Science
  Conference}\/}.
\newblock volume~\bibinfo{volume}{29} of {\it \bibinfo{series}{Lunar and
  Planetary Science Conference}\/}.
\bibitem[{{Neslusan} et~al.(1998){Neslusan}, {Svoren} \&
  {Porubcan}}]{neslusan1998}
\bibinfo{author}{{Neslusan}, L.}, \bibinfo{author}{{Svoren}, J.}, \&
  \bibinfo{author}{{Porubcan}, V.} (\bibinfo{year}{1998}).
\newblock \bibinfo{title}{{A computer program for calculation of a theoretical
  meteor-stream radiant}}.
\newblock {\it \bibinfo{journal}{\aap}\/},  {\it \bibinfo{volume}{331}\/},
  \bibinfo{pages}{411--413}.
\bibitem[{{Neukum} et~al.(2001){Neukum}, {Ivanov} \& {Hartmann}}]{neukum2001}
\bibinfo{author}{{Neukum}, G.}, \bibinfo{author}{{Ivanov}, B.~A.}, \&
  \bibinfo{author}{{Hartmann}, W.~K.} (\bibinfo{year}{2001}).
\newblock \bibinfo{title}{{Cratering Records in the Inner Solar System in
  Relation to the Lunar Reference System}}.
\newblock {\it \bibinfo{journal}{\ssr}\/},  {\it \bibinfo{volume}{96}\/},
  \bibinfo{pages}{55--86}.
\bibitem[{{Ortiz} et~al.(1999){Ortiz}, {Aceituno} \& {Aceituno}}]{ortiz1999}
\bibinfo{author}{{Ortiz}, J.~L.}, \bibinfo{author}{{Aceituno}, F.~J.}, \&
  \bibinfo{author}{{Aceituno}, J.} (\bibinfo{year}{1999}).
\newblock \bibinfo{title}{{A search for meteoritic flashes on the Moon}}.
\newblock {\it \bibinfo{journal}{\aap}\/},  {\it \bibinfo{volume}{343}\/},
  \bibinfo{pages}{L57--L60}.
\bibitem[{{Ortiz} et~al.(2006){Ortiz}, {Aceituno}, {Quesada}, {Aceituno},
  {Fern{\'a}ndez}, {Santos-Sanz}, {Trigo-Rodr{\'{\i}}guez}, {Llorca},
  {Mart{\'{\i}}n-Torres}, {Monta{\~n}{\'e}s-Rodr{\'{\i}}guez} \&
  {Pall{\'e}}}]{ortiz2006}
\bibinfo{author}{{Ortiz}, J.~L.}, \bibinfo{author}{{Aceituno}, F.~J.},
  \bibinfo{author}{{Quesada}, J.~A.}, \bibinfo{author}{{Aceituno}, J.},
  \bibinfo{author}{{Fern{\'a}ndez}, M.}, \bibinfo{author}{{Santos-Sanz}, P.},
  \bibinfo{author}{{Trigo-Rodr{\'{\i}}guez}, J.~M.}, \bibinfo{author}{{Llorca},
  J.}, \bibinfo{author}{{Mart{\'{\i}}n-Torres}, F.~J.},
  \bibinfo{author}{{Monta{\~n}{\'e}s-Rodr{\'{\i}}guez}, P.}, \&
  \bibinfo{author}{{Pall{\'e}}, E.} (\bibinfo{year}{2006}).
\newblock \bibinfo{title}{{Detection of sporadic impact flashes on the Moon:
  Implications for the luminous efficiency of hypervelocity impacts and derived
  terrestrial impact rates}}.
\newblock {\it \bibinfo{journal}{Icarus}\/},  {\it \bibinfo{volume}{184}\/},
  \bibinfo{pages}{319--326}. \DOIprefix\doi{10.1016/j.icarus.2006.05.002}.
\bibitem[{{Ortiz} et~al.(2015){Ortiz}, {Madiedo}, {Morales}, {Santos-Sanz} \&
  {Aceituno}}]{ortiz2015}
\bibinfo{author}{{Ortiz}, J.~L.}, \bibinfo{author}{{Madiedo}, J.~M.},
  \bibinfo{author}{{Morales}, N.}, \bibinfo{author}{{Santos-Sanz}, P.}, \&
  \bibinfo{author}{{Aceituno}, F.~J.} (\bibinfo{year}{2015}).
\newblock \bibinfo{title}{{Lunar impact flashes from Geminids: analysis of
  luminous efficiencies and the flux of large meteoroids on Earth}}.
\newblock {\it \bibinfo{journal}{\mnras}\/},  {\it \bibinfo{volume}{454}\/},
  \bibinfo{pages}{344--352}. \DOIprefix\doi{10.1093/mnras/stv1921}.
  \href{http://arxiv.org/abs/1511.07153}{\tt arXiv:1511.07153}.
\bibitem[{{Ortiz} et~al.(2002){Ortiz}, {Quesada}, {Aceituno}, {Aceituno} \&
  {Bellot Rubio}}]{ortiz2002}
\bibinfo{author}{{Ortiz}, J.~L.}, \bibinfo{author}{{Quesada}, J.~A.},
  \bibinfo{author}{{Aceituno}, J.}, \bibinfo{author}{{Aceituno}, F.~J.}, \&
  \bibinfo{author}{{Bellot Rubio}, L.~R.} (\bibinfo{year}{2002}).
\newblock \bibinfo{title}{{Observation and Interpretation of Leonid Impact
  Flashes on the Moon in 2001}}.
\newblock {\it \bibinfo{journal}{\apj}\/},  {\it \bibinfo{volume}{576}\/},
  \bibinfo{pages}{567--573}. \DOIprefix\doi{10.1086/341625}.
\bibitem[{{Ortiz} et~al.(2000){Ortiz}, {Sada}, {Bellot Rubio}, {Aceituno},
  {Aceituno}, {Guti{\'e}rrez} \& {Thiele}}]{ortiz2000}
\bibinfo{author}{{Ortiz}, J.~L.}, \bibinfo{author}{{Sada}, P.~V.},
  \bibinfo{author}{{Bellot Rubio}, L.~R.}, \bibinfo{author}{{Aceituno}, F.~J.},
  \bibinfo{author}{{Aceituno}, J.}, \bibinfo{author}{{Guti{\'e}rrez}, P.~J.},
  \& \bibinfo{author}{{Thiele}, U.} (\bibinfo{year}{2000}).
\newblock \bibinfo{title}{{Optical detection of meteoroidal impacts on the
  Moon}}.
\newblock {\it \bibinfo{journal}{\nat}\/},  {\it \bibinfo{volume}{405}\/},
  \bibinfo{pages}{921--923}.
\bibitem[{{Plescia} et~al.(2016){Plescia}, {Robinson}, {Wagner} \&
  {Baldridge}}]{plescia2016}
\bibinfo{author}{{Plescia}, J.~B.}, \bibinfo{author}{{Robinson}, M.~S.},
  \bibinfo{author}{{Wagner}, R.}, \& \bibinfo{author}{{Baldridge}, R.}
  (\bibinfo{year}{2016}).
\newblock \bibinfo{title}{{Ranger and Apollo S-IVB spacecraft impact craters}}.
\newblock {\it \bibinfo{journal}{\planss}\/},  {\it \bibinfo{volume}{124}\/},
  \bibinfo{pages}{15--35}. \DOIprefix\doi{10.1016/j.pss.2016.01.002}.
\bibitem[{{Robinson} et~al.(2015){Robinson}, {Boyd}, {Denevi}, {Lawrence},
  {McEwen}, {Moser}, {Povilaitis}, {Stelling}, {Suggs}, {Thompson} \&
  {Wagner}}]{robinson2015}
\bibinfo{author}{{Robinson}, M.~S.}, \bibinfo{author}{{Boyd}, A.~K.},
  \bibinfo{author}{{Denevi}, B.~W.}, \bibinfo{author}{{Lawrence}, S.~J.},
  \bibinfo{author}{{McEwen}, A.~S.}, \bibinfo{author}{{Moser}, D.~E.},
  \bibinfo{author}{{Povilaitis}, R.~Z.}, \bibinfo{author}{{Stelling}, R.~W.},
  \bibinfo{author}{{Suggs}, R.~M.}, \bibinfo{author}{{Thompson}, S.~D.}, \&
  \bibinfo{author}{{Wagner}, R.~V.} (\bibinfo{year}{2015}).
\newblock \bibinfo{title}{{New crater on the Moon and a swarm of secondaries}}.
\newblock {\it \bibinfo{journal}{\icarus}\/},  {\it \bibinfo{volume}{252}\/},
  \bibinfo{pages}{229--235}. \DOIprefix\doi{10.1016/j.icarus.2015.01.019}.
\bibitem[{{Schmidt} \& {Housen}(1987)}]{housen1987}
\bibinfo{author}{{Schmidt}, R.~M.}, \& \bibinfo{author}{{Housen}, K.~R.}
  (\bibinfo{year}{1987}).
\newblock \bibinfo{title}{{Some recent advances in the scaling of impact and
  explosion cratering}}.
\newblock {\it \bibinfo{journal}{International Journal of Impact
  Engineering}\/},  {\it \bibinfo{volume}{5}\/}, \bibinfo{pages}{543--560}.
\bibitem[{{Schultz} et~al.(2007){Schultz}, {Eberhardy}, {Ernst}, {A'Hearn},
  {Sunshine} \& {Lisse}}]{schultz2007}
\bibinfo{author}{{Schultz}, P.~H.}, \bibinfo{author}{{Eberhardy}, C.~A.},
  \bibinfo{author}{{Ernst}, C.~M.}, \bibinfo{author}{{A'Hearn}, M.~F.},
  \bibinfo{author}{{Sunshine}, J.~M.}, \& \bibinfo{author}{{Lisse}, C.~M.}
  (\bibinfo{year}{2007}).
\newblock \bibinfo{title}{{The Deep Impact oblique impact cratering
  experiment}}.
\newblock {\it \bibinfo{journal}{\icarus}\/},  {\it \bibinfo{volume}{191}\/},
  \bibinfo{pages}{84--122}. \DOIprefix\doi{10.1016/j.icarus.2007.06.031}.
\bibitem[{{Schultz} et~al.(2005){Schultz}, {Ernst} \& {Anderson}}]{schultz2005}
\bibinfo{author}{{Schultz}, P.~H.}, \bibinfo{author}{{Ernst}, C.~M.}, \&
  \bibinfo{author}{{Anderson}, J.~L.~B.} (\bibinfo{year}{2005}).
\newblock \bibinfo{title}{{Expectations for Crater Size and Photometric
  Evolution from the Deep Impact Collision}}.
\newblock {\it \bibinfo{journal}{SSR}\/},  {\it \bibinfo{volume}{117}\/},
  \bibinfo{pages}{207}. \DOIprefix\doi{10.1007/s11214-005-3383-7}.
\bibitem[{{Schultz} et~al.(2010){Schultz}, {Hermalyn}, {Colaprete}, {Ennico},
  {Shirley} \& {Marshall}}]{schultz2010}
\bibinfo{author}{{Schultz}, P.~H.}, \bibinfo{author}{{Hermalyn}, B.},
  \bibinfo{author}{{Colaprete}, A.}, \bibinfo{author}{{Ennico}, K.},
  \bibinfo{author}{{Shirley}, M.}, \& \bibinfo{author}{{Marshall}, W.~S.}
  (\bibinfo{year}{2010}).
\newblock \bibinfo{title}{{The LCROSS Cratering Experiment}}.
\newblock {\it \bibinfo{journal}{Science}\/},  {\it \bibinfo{volume}{330}\/},
  \bibinfo{pages}{468}. \DOIprefix\doi{10.1126/science.1187454}.
\bibitem[{{Schultz} et~al.(2013){Schultz}, {Hermalyn} \&
  {Veverka}}]{schultz2013}
\bibinfo{author}{{Schultz}, P.~H.}, \bibinfo{author}{{Hermalyn}, B.}, \&
  \bibinfo{author}{{Veverka}, J.} (\bibinfo{year}{2013}).
\newblock \bibinfo{title}{{The Deep Impact crater on 9P/Tempel-1 from
  Stardust-NExT}}.
\newblock {\it \bibinfo{journal}{\icarus}\/},  {\it \bibinfo{volume}{222}\/},
  \bibinfo{pages}{502--515}. \DOIprefix\doi{10.1016/j.icarus.2012.06.018}.
\bibitem[{{Sheward} et~al.(2019){Sheward}, {Avdellidou} \&
  {Sefton-Nash}}]{sheward2019}
\bibinfo{author}{{Sheward}, D.}, \bibinfo{author}{{Avdellidou}, C.}, \&
  \bibinfo{author}{{Sefton-Nash}, E.} (\bibinfo{year}{2019}).
\newblock \bibinfo{title}{{PyNAPLE: Automated Lunar Impact Flash Crater
  Detection}}.
\newblock In {\it \bibinfo{booktitle}{EPSC-DPS Joint Meeting 2019}\/} (pp.
  \bibinfo{pages}{EPSC--DPS2019--1032}).
\newblock volume \bibinfo{volume}{2019}.
\bibitem[{{Singer} et~al.(2019){Singer}, {McKinnon}, {Gladman}, {Greenstreet},
  {Bierhaus}, {Stern}, {Parker}, {Robbins}, {Schenk}, {Grundy}, {Bray},
  {Beyer}, {Binzel}, {Weaver}, {Young}, {Spencer}, {Kavelaars}, {Moore},
  {Zangari}, {Olkin}, {Lauer}, {Lisse}, {Ennico}, {New Horizons Geology}, Team,
  {New Horizons Surface Composition Science Theme Team} \& {New Horizons Ralph
  and LORRI Teams}}]{singer2019}
\bibinfo{author}{{Singer}, K.~N.}, \bibinfo{author}{{McKinnon}, W.~B.},
  \bibinfo{author}{{Gladman}, B.}, \bibinfo{author}{{Greenstreet}, S.},
  \bibinfo{author}{{Bierhaus}, E.~B.}, \bibinfo{author}{{Stern}, S.~A.},
  \bibinfo{author}{{Parker}, A.~H.}, \bibinfo{author}{{Robbins}, S.~J.},
  \bibinfo{author}{{Schenk}, P.~M.}, \bibinfo{author}{{Grundy}, W.~M.},
  \bibinfo{author}{{Bray}, V.~J.}, \bibinfo{author}{{Beyer}, R.~A.},
  \bibinfo{author}{{Binzel}, R.~P.}, \bibinfo{author}{{Weaver}, H.~A.},
  \bibinfo{author}{{Young}, L.~A.}, \bibinfo{author}{{Spencer}, J.~R.},
  \bibinfo{author}{{Kavelaars}, J.~J.}, \bibinfo{author}{{Moore}, J.~M.},
  \bibinfo{author}{{Zangari}, A.~M.}, \bibinfo{author}{{Olkin}, C.~B.},
  \bibinfo{author}{{Lauer}, T.~R.}, \bibinfo{author}{{Lisse}, C.~M.},
  \bibinfo{author}{{Ennico}, K.}, \bibinfo{author}{{New Horizons Geology}, G.},
  \bibinfo{author}{Team, I. S.~T.}, \bibinfo{author}{{New Horizons Surface
  Composition Science Theme Team}}, \& \bibinfo{author}{{New Horizons Ralph and
  LORRI Teams}} (\bibinfo{year}{2019}).
\newblock \bibinfo{title}{{Impact craters on Pluto and Charon indicate a
  deficit of small Kuiper belt objects}}.
\newblock {\it \bibinfo{journal}{Science}\/},  {\it \bibinfo{volume}{363}\/},
  \bibinfo{pages}{955--959}. \DOIprefix\doi{10.1126/science.aap8628}.
  \href{http://arxiv.org/abs/1902.10795}{\tt arXiv:1902.10795}.
\bibitem[{{Steyaert}(1991)}]{steyaert1991}
\bibinfo{author}{{Steyaert}, C.} (\bibinfo{year}{1991}).
\newblock \bibinfo{title}{{Calculating the solar longitude 2000.0.}}
\newblock {\it \bibinfo{journal}{WGN, Journal of the International Meteor
  Organization}\/},  {\it \bibinfo{volume}{19}\/}, \bibinfo{pages}{31--34}.
\bibitem[{{Stooke}(2019)}]{stooke2019}
\bibinfo{author}{{Stooke}, P.~J.} (\bibinfo{year}{2019}).
\newblock \bibinfo{title}{{Identification of the SMART-1 spacecraft impact
  location on the Moon}}.
\newblock {\it \bibinfo{journal}{\icarus}\/},  {\it \bibinfo{volume}{321}\/},
  \bibinfo{pages}{112--115}. \DOIprefix\doi{10.1016/j.icarus.2018.11.009}.
\bibitem[{{Strycker} et~al.(2013){Strycker}, {Chanover}, {Miller}, {Hamilton},
  {Hermalyn}, {Suggs} \& {Sussman}}]{strycker2013}
\bibinfo{author}{{Strycker}, P.~D.}, \bibinfo{author}{{Chanover}, N.~J.},
  \bibinfo{author}{{Miller}, C.}, \bibinfo{author}{{Hamilton}, R.~T.},
  \bibinfo{author}{{Hermalyn}, B.}, \bibinfo{author}{{Suggs}, R.~M.}, \&
  \bibinfo{author}{{Sussman}, M.} (\bibinfo{year}{2013}).
\newblock \bibinfo{title}{{Characterization of the LCROSS impact plume from a
  ground-based imaging detection}}.
\newblock {\it \bibinfo{journal}{Nature Communications}\/},  {\it
  \bibinfo{volume}{4}\/}, \bibinfo{pages}{2620}.
  \DOIprefix\doi{10.1038/ncomms3620}.
\bibitem[{{Suggs} et~al.(2014){Suggs}, {Moser}, {Cooke} \& {Suggs}}]{suggs2014}
\bibinfo{author}{{Suggs}, R.~M.}, \bibinfo{author}{{Moser}, D.~E.},
  \bibinfo{author}{{Cooke}, W.~J.}, \& \bibinfo{author}{{Suggs}, R.~J.}
  (\bibinfo{year}{2014}).
\newblock \bibinfo{title}{{The flux of kilogram-sized meteoroids from lunar
  impact monitoring}}.
\newblock {\it \bibinfo{journal}{Icarus}\/},  {\it \bibinfo{volume}{238}\/},
  \bibinfo{pages}{23--36}. \DOIprefix\doi{10.1016/j.icarus.2014.04.032}.
  \href{http://arxiv.org/abs/1404.6458}{\tt arXiv:1404.6458}.
\bibitem[{{Sugita} et~al.(2019){Sugita}, {Honda}, {Morota}, {Kameda}, {Sawada},
  {Tatsumi}, {Yamada}, {Honda}, {Yokota}, {Kouyama}, {Sakatani}, {Ogawa},
  {Suzuki}, {Okada}, {Namiki}, {Tanaka}, {Iijima}, {Yoshioka}, {Hayakawa},
  {Cho}, {Matsuoka}, {Hirata}, {Hirata}, {Miyamoto}, {Domingue}, {Hirabayashi},
  {Nakamura}, {Hiroi}, {Michikami}, {Michel}, {Ballouz}, {Barnouin}, {Ernst},
  {Schr{\"o}der}, {Kikuchi}, {Hemmi}, {Komatsu}, {Fukuhara}, {Taguchi}, {Arai},
  {Senshu}, {Demura}, {Ogawa}, {Shimaki}, {Sekiguchi}, {M{\"u}ller},
  {Hagermann}, {Mizuno}, {Noda}, {Matsumoto}, {Yamada}, {Ishihara}, {Ikeda},
  {Araki}, {Yamamoto}, {Abe}, {Yoshida}, {Higuchi}, {Sasaki}, {Oshigami},
  {Tsuruta}, {Asari}, {Tazawa}, {Shizugami}, {Kimura}, {Otsubo}, {Yabuta},
  {Hasegawa}, {Ishiguro}, {Tachibana}, {Palmer}, {Gaskell}, {Le Corre},
  {Jaumann}, {Otto}, {Schmitz}, {Abell}, {Barucci}, {Zolensky}, {Vilas},
  {Thuillet}, {Sugimoto}, {Takaki}, {Suzuki}, {Kamiyoshihara}, {Okada},
  {Nagata}, {Fujimoto}, {Yoshikawa}, {Yamamoto}, {Shirai}, {Noguchi}, {Ogawa},
  {Terui}, {Kikuchi}, {Yamaguchi}, {Oki}, {Takao}, {Takeuchi}
  et~al.}]{sugita2019}
\bibinfo{author}{{Sugita}, S.}, \bibinfo{author}{{Honda}, R.},
  \bibinfo{author}{{Morota}, T.}, \bibinfo{author}{{Kameda}, S.},
  \bibinfo{author}{{Sawada}, H.}, \bibinfo{author}{{Tatsumi}, E.},
  \bibinfo{author}{{Yamada}, M.}, \bibinfo{author}{{Honda}, C.},
  \bibinfo{author}{{Yokota}, Y.}, \bibinfo{author}{{Kouyama}, T.},
  \bibinfo{author}{{Sakatani}, N.}, \bibinfo{author}{{Ogawa}, K.},
  \bibinfo{author}{{Suzuki}, H.}, \bibinfo{author}{{Okada}, T.},
  \bibinfo{author}{{Namiki}, N.}, \bibinfo{author}{{Tanaka}, S.},
  \bibinfo{author}{{Iijima}, Y.}, \bibinfo{author}{{Yoshioka}, K.},
  \bibinfo{author}{{Hayakawa}, M.}, \bibinfo{author}{{Cho}, Y.},
  \bibinfo{author}{{Matsuoka}, M.}, \bibinfo{author}{{Hirata}, N.},
  \bibinfo{author}{{Hirata}, N.}, \bibinfo{author}{{Miyamoto}, H.},
  \bibinfo{author}{{Domingue}, D.}, \bibinfo{author}{{Hirabayashi}, M.},
  \bibinfo{author}{{Nakamura}, T.}, \bibinfo{author}{{Hiroi}, T.},
  \bibinfo{author}{{Michikami}, T.}, \bibinfo{author}{{Michel}, P.},
  \bibinfo{author}{{Ballouz}, R.~L.}, \bibinfo{author}{{Barnouin}, O.~S.},
  \bibinfo{author}{{Ernst}, C.~M.}, \bibinfo{author}{{Schr{\"o}der}, S.~E.},
  \bibinfo{author}{{Kikuchi}, H.}, \bibinfo{author}{{Hemmi}, R.},
  \bibinfo{author}{{Komatsu}, G.}, \bibinfo{author}{{Fukuhara}, T.},
  \bibinfo{author}{{Taguchi}, M.}, \bibinfo{author}{{Arai}, T.},
  \bibinfo{author}{{Senshu}, H.}, \bibinfo{author}{{Demura}, H.},
  \bibinfo{author}{{Ogawa}, Y.}, \bibinfo{author}{{Shimaki}, Y.},
  \bibinfo{author}{{Sekiguchi}, T.}, \bibinfo{author}{{M{\"u}ller}, T.~G.},
  \bibinfo{author}{{Hagermann}, A.}, \bibinfo{author}{{Mizuno}, T.},
  \bibinfo{author}{{Noda}, H.}, \bibinfo{author}{{Matsumoto}, K.},
  \bibinfo{author}{{Yamada}, R.}, \bibinfo{author}{{Ishihara}, Y.},
  \bibinfo{author}{{Ikeda}, H.}, \bibinfo{author}{{Araki}, H.},
  \bibinfo{author}{{Yamamoto}, K.}, \bibinfo{author}{{Abe}, S.},
  \bibinfo{author}{{Yoshida}, F.}, \bibinfo{author}{{Higuchi}, A.},
  \bibinfo{author}{{Sasaki}, S.}, \bibinfo{author}{{Oshigami}, S.},
  \bibinfo{author}{{Tsuruta}, S.}, \bibinfo{author}{{Asari}, K.},
  \bibinfo{author}{{Tazawa}, S.}, \bibinfo{author}{{Shizugami}, M.},
  \bibinfo{author}{{Kimura}, J.}, \bibinfo{author}{{Otsubo}, T.},
  \bibinfo{author}{{Yabuta}, H.}, \bibinfo{author}{{Hasegawa}, S.},
  \bibinfo{author}{{Ishiguro}, M.}, \bibinfo{author}{{Tachibana}, S.},
  \bibinfo{author}{{Palmer}, E.}, \bibinfo{author}{{Gaskell}, R.},
  \bibinfo{author}{{Le Corre}, L.}, \bibinfo{author}{{Jaumann}, R.},
  \bibinfo{author}{{Otto}, K.}, \bibinfo{author}{{Schmitz}, N.},
  \bibinfo{author}{{Abell}, P.~A.}, \bibinfo{author}{{Barucci}, M.~A.},
  \bibinfo{author}{{Zolensky}, M.~E.}, \bibinfo{author}{{Vilas}, F.},
  \bibinfo{author}{{Thuillet}, F.}, \bibinfo{author}{{Sugimoto}, C.},
  \bibinfo{author}{{Takaki}, N.}, \bibinfo{author}{{Suzuki}, Y.},
  \bibinfo{author}{{Kamiyoshihara}, H.}, \bibinfo{author}{{Okada}, M.},
  \bibinfo{author}{{Nagata}, K.}, \bibinfo{author}{{Fujimoto}, M.},
  \bibinfo{author}{{Yoshikawa}, M.}, \bibinfo{author}{{Yamamoto}, Y.},
  \bibinfo{author}{{Shirai}, K.}, \bibinfo{author}{{Noguchi}, R.},
  \bibinfo{author}{{Ogawa}, N.}, \bibinfo{author}{{Terui}, F.},
  \bibinfo{author}{{Kikuchi}, S.}, \bibinfo{author}{{Yamaguchi}, T.},
  \bibinfo{author}{{Oki}, Y.}, \bibinfo{author}{{Takao}, Y.},
  \bibinfo{author}{{Takeuchi}, H.} et~al. (\bibinfo{year}{2019}).
\newblock \bibinfo{title}{{The geomorphology, color, and thermal properties of
  Ryugu: Implications for parent-body processes}}.
\newblock {\it \bibinfo{journal}{Science}\/},  {\it \bibinfo{volume}{364}\/},
  \bibinfo{pages}{252--252}. \DOIprefix\doi{10.1126/science.aaw0422}.
\bibitem[{{Vincent} et~al.(2015){Vincent}, {Oklay}, {Marchi}, {H{\"o}fner} \&
  {Sierks}}]{vincent2015}
\bibinfo{author}{{Vincent}, J.-B.}, \bibinfo{author}{{Oklay}, N.},
  \bibinfo{author}{{Marchi}, S.}, \bibinfo{author}{{H{\"o}fner}, S.}, \&
  \bibinfo{author}{{Sierks}, H.} (\bibinfo{year}{2015}).
\newblock \bibinfo{title}{{Craters on comets}}.
\newblock {\it \bibinfo{journal}{\planss}\/},  {\it \bibinfo{volume}{107}\/},
  \bibinfo{pages}{53--63}. \DOIprefix\doi{10.1016/j.pss.2014.06.008}.
\bibitem[{Walsh et~al.(2019)Walsh, Jawin, Ballouz, Barnouin, Bierhaus,
  Connolly, Molaro, McCoy, Delbo, Hartzell, Pajola, Schwartz, Trang, Asphaug,
  Becker, Beddingfield, Bennett, Bottke, Burke, Clark, Daly, DellaGiustina,
  Dworkin, Elder, Golish, Hildebrand, Malhotra, Marshall, Michel, Nolan, Perry,
  Rizk, Ryan, Sandford, Scheeres, Susorney, Thuillet, Lauretta \&
  Team}]{walsh2019}
\bibinfo{author}{Walsh, K.~J.}, \bibinfo{author}{Jawin, E.~R.},
  \bibinfo{author}{Ballouz, R.~L.}, \bibinfo{author}{Barnouin, O.~S.},
  \bibinfo{author}{Bierhaus, E.~B.}, \bibinfo{author}{Connolly, H.~C.},
  \bibinfo{author}{Molaro, J.~L.}, \bibinfo{author}{McCoy, T.~J.},
  \bibinfo{author}{Delbo, M.}, \bibinfo{author}{Hartzell, C.~M.},
  \bibinfo{author}{Pajola, M.}, \bibinfo{author}{Schwartz, S.~R.},
  \bibinfo{author}{Trang, D.}, \bibinfo{author}{Asphaug, E.},
  \bibinfo{author}{Becker, K.~J.}, \bibinfo{author}{Beddingfield, C.~B.},
  \bibinfo{author}{Bennett, C.~A.}, \bibinfo{author}{Bottke, W.~F.},
  \bibinfo{author}{Burke, K.~N.}, \bibinfo{author}{Clark, B.~C.},
  \bibinfo{author}{Daly, M.~G.}, \bibinfo{author}{DellaGiustina, D.~N.},
  \bibinfo{author}{Dworkin, J.~P.}, \bibinfo{author}{Elder, C.~M.},
  \bibinfo{author}{Golish, D.~R.}, \bibinfo{author}{Hildebrand, A.~R.},
  \bibinfo{author}{Malhotra, R.}, \bibinfo{author}{Marshall, J.},
  \bibinfo{author}{Michel, P.}, \bibinfo{author}{Nolan, M.~C.},
  \bibinfo{author}{Perry, M.~E.}, \bibinfo{author}{Rizk, B.},
  \bibinfo{author}{Ryan, A.}, \bibinfo{author}{Sandford, S.~A.},
  \bibinfo{author}{Scheeres, D.~J.}, \bibinfo{author}{Susorney, H. C.~M.},
  \bibinfo{author}{Thuillet, F.}, \bibinfo{author}{Lauretta, D.~S.}, \&
  \bibinfo{author}{Team, O.-R.} (\bibinfo{year}{2019}).
\newblock \bibinfo{title}{{Craters, boulders and regolith of (101955) Bennu
  indicative of an old and dynamic surface}}.
\newblock {\it \bibinfo{journal}{Nature Geoscience}\/},  {\it
  \bibinfo{volume}{12}\/}, \bibinfo{pages}{242--246}.
\bibitem[{{Xilouris} et~al.(2018){Xilouris}, {Bonanos}, {Bellas-Velidis},
  {Boumis}, {Dapergolas}, {Maroussis}, {Liakos}, {Alikakos}, {Charmandaris},
  {Dimou}, {Fytsilis}, {Kelley}, {Koschny}, {Navarro}, {Tsiganis} \&
  {Tsinganos}}]{xilouris2018}
\bibinfo{author}{{Xilouris}, E.~M.}, \bibinfo{author}{{Bonanos}, A.~Z.},
  \bibinfo{author}{{Bellas-Velidis}, I.}, \bibinfo{author}{{Boumis}, P.},
  \bibinfo{author}{{Dapergolas}, A.}, \bibinfo{author}{{Maroussis}, A.},
  \bibinfo{author}{{Liakos}, A.}, \bibinfo{author}{{Alikakos}, I.},
  \bibinfo{author}{{Charmandaris}, V.}, \bibinfo{author}{{Dimou}, G.},
  \bibinfo{author}{{Fytsilis}, A.}, \bibinfo{author}{{Kelley}, M.},
  \bibinfo{author}{{Koschny}, D.}, \bibinfo{author}{{Navarro}, V.},
  \bibinfo{author}{{Tsiganis}, K.}, \& \bibinfo{author}{{Tsinganos}, K.}
  (\bibinfo{year}{2018}).
\newblock \bibinfo{title}{{NELIOTA: The wide-field, high-cadence, lunar
  monitoring system at the prime focus of the Kryoneri telescope}}.
\newblock {\it \bibinfo{journal}{\aap}\/},  {\it \bibinfo{volume}{619}\/},
  \bibinfo{pages}{A141}. \DOIprefix\doi{10.1051/0004-6361/201833499}.
  \href{http://arxiv.org/abs/1809.00495}{\tt arXiv:1809.00495}.
\bibitem[{{Yanagisawa} \& {Kisaichi}(2002)}]{yanagisawa2002}
\bibinfo{author}{{Yanagisawa}, M.}, \& \bibinfo{author}{{Kisaichi}, N.}
  (\bibinfo{year}{2002}).
\newblock \bibinfo{title}{{Lightcurves of 1999 Leonid Impact Flashes on the
  Moon}}.
\newblock {\it \bibinfo{journal}{\icarus}\/},  {\it \bibinfo{volume}{159}\/},
  \bibinfo{pages}{31--38}. \DOIprefix\doi{10.1006/icar.2002.6931}.
\bibitem[{{Yanagisawa} et~al.(2006){Yanagisawa}, {Ohnishi}, {Takamura},
  {Masuda}, {Sakai}, {Ida}, {Adachi} \& {Ishida}}]{yanagisawa2006}
\bibinfo{author}{{Yanagisawa}, M.}, \bibinfo{author}{{Ohnishi}, K.},
  \bibinfo{author}{{Takamura}, Y.}, \bibinfo{author}{{Masuda}, H.},
  \bibinfo{author}{{Sakai}, Y.}, \bibinfo{author}{{Ida}, M.},
  \bibinfo{author}{{Adachi}, M.}, \& \bibinfo{author}{{Ishida}, M.}
  (\bibinfo{year}{2006}).
\newblock \bibinfo{title}{{The first confirmed Perseid lunar impact flash}}.
\newblock {\it \bibinfo{journal}{\icarus}\/},  {\it \bibinfo{volume}{182}\/},
  \bibinfo{pages}{489--495}. \DOIprefix\doi{10.1016/j.icarus.2006.01.004}.

\end{thebibliography}

\end{document}